\documentclass{article}
\usepackage{amsmath}
\usepackage{amssymb}
\usepackage{listings}
\usepackage{xcolor}
\usepackage{graphicx}
\usepackage{booktabs}
\usepackage{bytefield}
\usepackage[T1]{fontenc}
\usepackage[utf8]{inputenc}
\usepackage{subcaption}
\usepackage[margin=1in]{geometry}
\begin{document}

\title{A High-Throughput Compute-Efficient POMDP Hide-And-Seek-Engine (HASE) for Multi-Agent Operations}
\author{Timothy Flavin, Sandip Sen}
\date{4/29/2026}
\maketitle
\enlargethispage{\baselineskip}
\enlargethispage{\baselineskip}
\tableofcontents
\newpage

\begin{abstract}
Reinforcement Learning (RL) algorithms exhibit high sample complexity, particularly when applied to Decentralized Partially Observable Markov Decision Processes (Dec-POMDPs). As a response, projects such as SampleFactory, EnvPool, Brax, and IsaacLab migrate parallel execution of classic environments such as MuJoCo and Atari into C++ thread pools or the GPU to decrease the computational cost of environment steps. We are interested in optimizing the decision-level of human-AI joint operations, so we introduce a compute-efficient Dec-POMDP engine natively architected in C++ called Hide-And-Seek-Engine. By employing Data-Oriented Design (DOD) principles, explicit 64-byte cache-line alignment to remove false sharing, and a zero-copy PyTorch memory bridge using pinned memory and Direct Memory Access (DMA), our engine sustains throughput of up to 33,000,000 steps per second (SPS) in a single-agent, 1024-environment, decentralized observations on an AMD Ryzen 9950X (16 cores). Ten agents reduces FPS to 7M SPS with generating random actions contributing 1/3rd the total runtime for reference. The engine achieves a throughput increase of approximately 3,500$\times$ over the baseline single threaded vectorized NumPy implementation and successfully trains cooperative multi-agent policies via PPO, DQN, and SAC in minutes, validating both its performance and generality.
\end{abstract}

\section{Introduction}
\label{sec:introduction}
The deployment of autonomous agents in complex, unstructured environments such as Search and Rescue (SAR), ecological surveying, and military interdiction demands robust decision-making paradigms. These scenarios require dynamic handling of spatial partial observability, heterogeneous agent capabilities, and flexible mission parameters. In order to learn approximately optimal policies, Deep Multi-Agent Reinforcement Learning (MARL)~\cite{MARLreview} has emerged as a highly capable framework to resolve these Partially Observable Markov Decision Processes (POMDPs)~\cite{vinyals2019grandmaster,berner2019dota}. However, MARL algorithms such as Multi-Agent Proximal Policy Optimization (MAPPO), require high sample complexity~\cite{bernstein2002complexity, jin2020sample} to ensure stable policy updates. Online RL depends on continuous, high-volume influxes of fresh on-policy data to securely estimate gradients, sometimes accomplished asynchronously as in A3C~\cite{mnih2016asynchronous}. To stabilize training and prevent catastrophic forgetting, modern distributed frameworks enforce massive batch gathering across thousands of desynchronized environments and experience reuse~\cite{espeholt2018impala,apex,pdqn}. 

Scaling MARL to fulfill large sample throughput requirements is an active area of research~\cite{mittal2025isaac,cusumano2025robust,weng2022envpool,petrenko2020sample}. Python-based environments utilizing vectorized bulk math (e.g., NumPy) incur abstraction latency and are constrained by Python's Global Interpreter Lock (GIL) or multiprocessing library as in gymnasiums' VecEnv~\cite{towers2024gymnasium}. We empirically confirm in section \ref{sec:ablation} the costly overhead incurred by generic Python approaches. Even when ported to native C++ via OpenMP~\cite{openmp08}, there are a number of pitfalls when operating across many environments. Sparse one-hot agent-observation tensors cause frequent cache misses and small cross-thread arrays (such as ``rewards" or ``dones") cause thrashing from False Sharing~\cite{torrellas1994false}, a multi-threading penalty arising when independent CPU cores mutate distinct variables residing on the same physical cache line as one another. As we will show in sections (\ref{sec:high_performance},\ref{sec:hardware_tuning}), False sharing is only one of many performance bottlenecks guiding the data-oriented architectural decisions of Hide-And-Seek-Engine~\cite{arantes2025impact}. To overcome these bottlenecks, we present a high-performance, compute-efficient POMDP simulator for real-world multi-agent coordination. Through careful memory management, bit packing, and cache-line alignment, the engine substantially reduces cache misses and false sharing (Section~\ref{sec:high_performance}). A zero-copy PyTorch pinned memory bridge via PyBind11~\cite{pybind11}, and Direct Memory Access (DMA), and careful management of OMP thread pools to reduce starvation of PyTorch threads reduces CPU-to-GPU staging and transfer overhead (Section~\ref{sec:high_performance}). Together, these optimizations increase throughput from approximately 4,000 SPS (Gymnasium async baseline; see Section~\ref{sec:ablation}) to over 33,000,000 SPS in single-agent configurations on an AMD Ryzen 9950X, with less performance degradation as the number of agents grows. 

\section{Related Work}
\label{sec:related_work}
Standard MARL benchmarks such as the StarCraft Multi-Agent Challenge (SMAC)~\cite{samvelyan2019starcraft,ellis2023smacv2}, PettingZoo~\cite{terry2021pettingzoo}, and Unity ML-Agents~\cite{juliani2018unity} have driven significant algorithmic progress. However, their underlying simulation engines are often constrained by engine speed, Python's Global Interpreter Lock (GIL), or multiprocessing overhead~\cite{towers2024gymnasium}. To bypass CPU bounds, hardware-accelerated simulators like NVIDIA Isaac Gym~\cite{mittal2025isaac} and Google Brax~\cite{brax2021github} natively execute on GPUs. While highly performant for continuous control, GPU simulators experience degraded performance when evaluating the complex, sequential branching logic (e.g., discrete grid collisions, heterogeneous agent capabilities) present in many operational POMDPs due to warp-divergence~\cite{shacklett2023madrona}. As a result, there also exist a number of parallel environments and environment migrations towards C++ for execution on the CPU, including SampleFactory\cite{petrenko2020sample}, EnvPool~\cite{weng2022envpool}, Brax\cite{brax2021github}, PufferLib~\cite{suarez2024pufferlib}, IsaacGym~\cite{mittal2025isaac} and OpenSpiel~\cite{lanctot2019openspiel}. 
Functionally, our environment is most similar to MiniGrid~\cite{MinigridMiniworld23} in it's functionality as more of a grid-based game engine than a single environment. Unlike minigrid, movement is continuous, and the map is made to model multi-agent dynamics with dynamic in visibility, navigation speed, and targets at a much higher experiential throughput.  
%Furthermore, modern distributed RL architectures such as IMPALA~\cite{espeholt2018impala}, Ape-X~\cite{apex}, and A3C~\cite{mnih2016asynchronous} necessitate massive batching of desynchronized environments to stabilize training by acting on independent and identically distributed (I.I.D.) data, highlighting the need for highly scalable, cache-optimized CPU engines.

\section{Environment Operation and Functional Design}
\label{sec:env_operation}

\begin{figure}[htbp]
    \centering
    % Subfigure 1
    \begin{subfigure}[b]{0.24\textwidth}
        \centering
        \includegraphics[width=\textwidth]{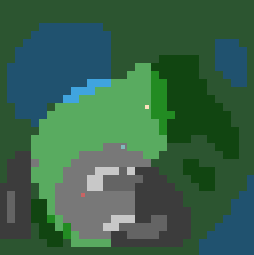}
        \caption{Mountain SAR}
        \label{fig:sar}
    \end{subfigure}
    \hfill
    % Subfigure 2
    \begin{subfigure}[b]{0.24\textwidth}
        \centering
        \includegraphics[width=\textwidth]{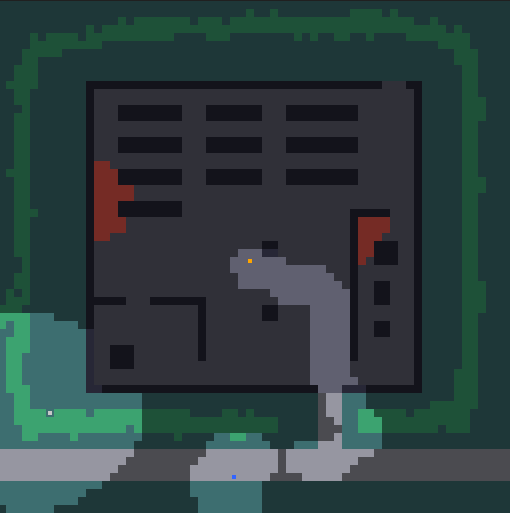}
        \caption{Warehouse Fire}
        \label{fig:warehouse}
    \end{subfigure}
    \hfill
    % Subfigure 3
    \begin{subfigure}[b]{0.24\textwidth}
        \centering
        \includegraphics[width=\textwidth]{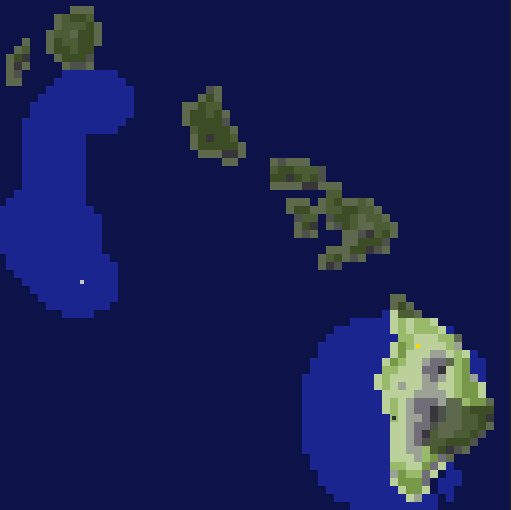}
        \caption{Coast Guard Monitoring}
        \label{fig:coastguard}
    \end{subfigure}
    \hfill
    % Subfigure 4
    \begin{subfigure}[b]{0.24\textwidth}
        \centering
        \includegraphics[width=\textwidth]{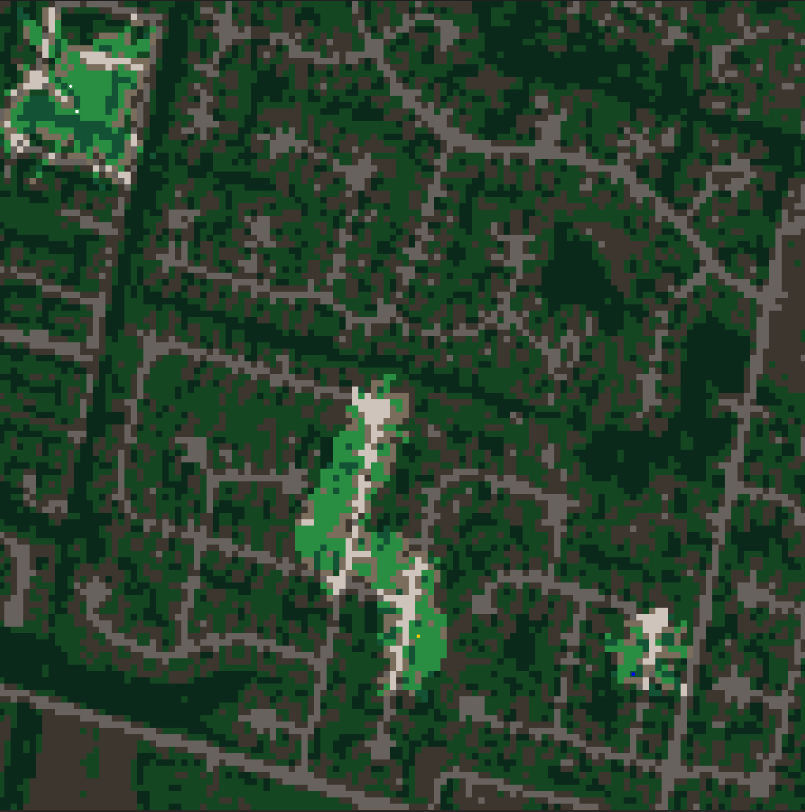}
        \caption{Neighborhood Watch}
        \label{fig:neighborhood}
    \end{subfigure}
    
    \caption{Overview of Emergency Response and Monitoring Scenarios (Single Row)}
    \label{fig:scenarios}
\end{figure}

Hide-And-Seek-Engine serves as a general-purpose testing ground for cooperative coverage, hide-and-seek contra-RL, and multi-agent navigation, where hiders either remain stationary (acting as waypoint targets) or can be controlled randomly or adversarially. The state space is exposed as an MDP, POMDP, and Dec-POMDP, supporting purely cooperative control over seekers or zero-sum adversarial control over both hiders and seekers. The engine provides a highly parallelized RL testing ground broadly applicable to any multi-agent search paradigm, such as SAR, illicit drug interdiction, or ecological surveying. The engine simulates up to 20 cooperating or independent searcher agents and an arbitrary number of hidden hider entities (referred to as Persons of Interest, or POIs). Observations consist of both spatial layers for Image processing and vectorized details for MLP processing; seeker locations reside on individual layers to differentiate heterogeneous teammate positions, and POI locations can occupy either unique one-hot-coded spatial layers or a single undifferentiated layer to scale arbitrarily.

\subsection{Formal Problem Formulation}
\label{sec:formal}
We formalize the scenario as a Dec-POMDP defined by the tuple $\langle \mathcal{I}, \mathcal{S}, \{\mathcal{A}_i\}_{i \in \mathcal{I}}, \mathcal{T}, \mathcal{R}, \{\Omega_i\}_{i \in \mathcal{I}}, \{\mathcal{O}_i\}_{i \in \mathcal{I}}, \gamma \rangle$, where $\mathcal{I}$ is the finite set of agents.
\begin{itemize}
    \item \textbf{State Space ($\mathcal{S}$):} The true environment state encompasses the global semantic grid, precise coordinates of all agents and Persons of Interest (POIs), and agent internal states (e.g., battery life, deployment status).
    \item \textbf{Action Space ($\mathcal{A}_i$):} Each agent operates within a Box2D action space governing directional movement and a discrete radio broadcast space designating targets to share information with.
    \item \textbf{Observation Space ($\Omega_i$):} Each agent's observation comprises a multi-dimensional tensor representing terrain data such dynamic altitudes and interactivity accross multiple semantic channels (e.g., blocking/non-blocking, observed/unknown, dangerous/traversable), coupled with a 1D logical vector containing the agent's key internal beliefs.
    \item \textbf{Observation Function ($\mathcal{O}_i$):} A deterministic line-of-sight projection from the true state $s \in \mathcal{S}$ to agent $i$'s observation $o_i \in \Omega_i$, conditioned on agent altitude.% and tile-level occlusion rules.
    \item \textbf{Reward Function ($\mathcal{R}$):} Agents receive positive semi-dense intrinsic rewards for exploring undiscovered tiles and locating POIs, with a sparse terminal reward acquired upon successfully ``saving" a POI.
\end{itemize}

\subsection{Configurable World Generation and Traversal}
\label{sec:world_gen}
To maximize generalizability across a wide range of scenario distributions, the engine is entirely data-driven. Map topologies are parsed at initialization from standard image files (e.g., PNG) alongside JSON configuration matrices. The engine converts RGB values into concrete semantic tile types via an exact-match lookup table mapping RGB triplets to predefined tile type identifiers. Each tile type governs specific environmental interactions:
\begin{itemize}
    \item \textbf{Traversal Kinetics:} Modifiers specifying whether a tile is Walkable, Aquatic, or Flyable, permitting heterogeneous agent configurations (e.g., quadcopters, boats, and foot-searchers operating synchronously).
    \item \textbf{Collision Models:} Tiles strictly define \textit{blocking} (impassable obstacles like walls) or \textit{non-blocking} behavior, which dictates the vector-based continuous movement mechanics of the agents such as move speed or probability of getting stuck. Stuck agents need to be rescued by a teammate to regain movement capability.
    \item \textbf{Altitude and View Ranges:} Dynamic, continuous terrain altitudes combine with inherent agent view ranges to handle dynamic visibility and flight suitability across levels. Grounded agents can see further at higher altitudes and flying agents cannot pass terrain above a certain altitude.
    \item \textbf{Agent-Tile Move-speed:} For each agent-tile pairing, a move-speed can be defined via the config files for that agent on that tile. This way individual agents can vary in their ratio of walking to swim speed, for example.
\end{itemize}

\subsection{Communication, Observability, and Vectorization}
\label{sec:comm}
The engine supports sophisticated multi-agent systems via tunable built-in radio settings. Agents can produce both movement and radio actions, broadcasting their localized beliefs to other individual agents within the native C++ observation representation. By choosing themselves as the target, agents share nothing over the radio so that communication can be handled externally by a learning algorithm instead. Shared information includes discovered POI locations, local terrain, and agent states, organically resolving asynchronous knowledge states. When an agent chooses to share information, it reveals all tiles it can currently see, and all information it has on the POI's to the agent it messages.

Furthermore, depending on the specific MARL algorithm being trained, researchers can independently configure the environment's returned observations (decentralized, centralized, or none) and access to the true game state (included or ignored). This combinatorial design results in six distinct observability modes:
\begin{enumerate}
    \item \textbf{Decentralized Observation (No State):} The standard Dec-POMDP setting for real-world deployment. Each agent receives a unique observation tensor restricted to its functional line-of-sight and local radio telemetry, and the true game state is withheld.
    \item \textbf{Decentralized Observation with True State:} Designed for Centralized Training with Decentralized Execution (CTDE). Agents receive localized observations for their actor networks, while the environment concurrently returns the omniscient true state to optimize the centralized critic.
    \item \textbf{Centralized Observation (No State):} Fuses all active agent sightlines into a singular, shared partial observation tensor, without exposing the underlying global game state.
    \item \textbf{Centralized Observation with True State:} Provides the fused, shared partial observation tensor alongside unrestricted access to the full, hidden game state.
    \item \textbf{True State Only (No Observation):} Agents receive no localized or centralized partial observations. The environment solely returns the omniscient game state, typically utilized for baseline debugging or fully centralized algorithmic controllers.
    \item \textbf{Void Output (No Observation, No State):} The environment bypasses both observation generation and state returns, yielding an empty output (used for native C++ integrations such as traditional search that may skip the expensive tensor fill and use the game state directly).
\end{enumerate}

To ensure robust and stable RL training, environment instances within a vectorized batch are stepped stochastically during initialization. This desynchronized execution horizon prevents episodic correlation spikes (where all environments in a batch terminate and reset simultaneously), thereby producing approximately i.i.d.\ samples within each training batch~\cite{mnih2015human}, stabilizing batch gradients and improving online algorithmic convergence.

\section{High-Performance C++ Architecture}
\label{sec:high_performance}

By constructing a cache-optimized C++ backend coupled with a zero-copy PyTorch memory bridge, our engine achieves throughputs exceeding 33,000,000 SPS in single-agent configurations on an AMD Ryzen 9950X and over 5,000,000 SPS on a low-mid range quad core laptop (see Appendix~\ref{app:hardware_runtimes}). This represents a throughput increase of approximately 3,000$\times$ compared to a NumPy baseline at approximately 2,000 SPS. This section details the architectural decisions that yield this speedup.

\subsection{Cache-Aligned Hierarchical Memory Arena}
\label{sec:arena}
\begin{figure}[ht]
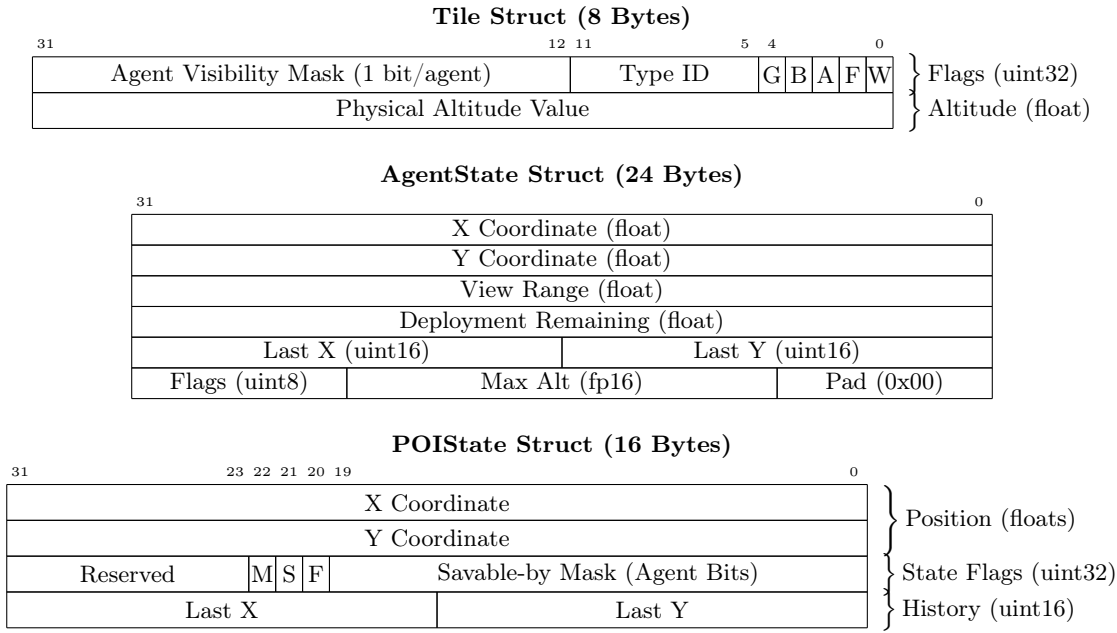

\title{Bit Packing Environment Structure}
    \centering
    \small
    \vspace{0.5em} % Significant space between different structs
    
    % --- TILE STRUCT ---
    \textbf{Tile Struct (8 Bytes)} \\[1.5ex] % Buffer between title and header
    \begin{bytefield}[endianness=big, bitwidth=1.1em, bitheight=3.5ex]{32}
        \bitheader{0,4,5,11,12,31} \\
        \begin{rightwordgroup}{Flags  (uint32)}
            \bitbox{20}{Agent Visibility Mask (1 bit/agent)} & 
            \bitbox{7}{Type ID} & 
            \bitbox{1}{G} & \bitbox{1}{B} & \bitbox{1}{A} & \bitbox{1}{F} & \bitbox{1}{W}
        \end{rightwordgroup} \\
        \begin{rightwordgroup}{Altitude (float)}
            \bitbox{32}{Physical Altitude Value}
        \end{rightwordgroup}
    \end{bytefield}

    \vspace{1em} % Significant space between different structs

    % --- AGENT STATE ---
    \textbf{AgentState Struct (24 Bytes)} \\[1.5ex]
    \begin{bytefield}[endianness=big, bitwidth=1.1em, bitheight=3ex]{32}
        \bitheader{0,31} \\
        \bitbox{32}{X Coordinate (float)} \\
        \bitbox{32}{Y Coordinate (float)} \\
        \bitbox{32}{View Range (float)} \\
        \bitbox{32}{Deployment Remaining (float)} \\
        \bitbox{16}{Last X (uint16)} & \bitbox{16}{Last Y (uint16)} \\
        \bitbox{8}{Flags (uint8)} & \bitbox{16}{Max Alt (fp16)} & \bitbox{8}{Pad (0x00)}
    \end{bytefield}

    \vspace{1em}

    % --- POI STATE ---
    \textbf{POIState Struct (16 Bytes)} \\[1.5ex]
    \begin{bytefield}[endianness=big, bitwidth=1.1em, bitheight=3.5ex]{32}
        \bitheader{0,19,20,21,22,23,31} \\
        \begin{rightwordgroup}{Position (floats)}
            \bitbox{32}{X Coordinate} \\
            \bitbox{32}{Y Coordinate}
        \end{rightwordgroup} \\
        \begin{rightwordgroup}{State Flags (uint32)}
            \bitbox{9}{Reserved} & 
            \bitbox{1}{M} & \bitbox{1}{S} & \bitbox{1}{F} & 
            \bitbox{20}{Savable-by Mask (Agent Bits)}
        \end{rightwordgroup} \\
        \begin{rightwordgroup}{History (uint16)}
            \bitbox{16}{Last X} & \bitbox{16}{Last Y}
        \end{rightwordgroup}
    \end{bytefield}

    \vspace{0.5em}
    \caption{Detailed memory layout and bit-packing for environment entities. Metadata is compressed into 32-bit words to maximize L1 cache residency and minimize memory bandwidth overhead. 
\textbf{Tile Flags:} \textbf{G}~--~Global observed status, \textbf{B}~--~Blocking/Impassable, \textbf{A}~--~Aquatic, \textbf{F}~--~Flyable, and \textbf{W}~--~Walkable. 
\textbf{AgentState Flags:} \textbf{bit 0}~--~Stuck, \textbf{bit 1}~--~Walk, \textbf{bit 2}~--~Fly, \textbf{bit 3}~--~Swim.
\textbf{POIState Flags:} \textbf{M}~--~Moves (dynamic entity), \textbf{S}~--~Saved/Rescued, and \textbf{F}~--~Found/Discovered. 
The 20-bit masks allow for $O(1)$ visibility and capability checks via bitwise AND/OR operations across up to 20 agents.}
    \label{fig:env_bytefields_fixed}
\end{figure}

At the core of the engine's performance is the \texttt{EnvironmentArena} (Figure~\ref{fig:env_bytefields_fixed}, table \ref{fig:internal_mem_layout}, language adopted data oriented design principals~\cite{arantes2025impact}), which replaces standard heap-allocated object trees with a localized, contiguous flat-memory slab (\texttt{std::vector<uint8\_t> memory}). Grid-based MARL environments typically suffer from severe cache fragmentation as agents traverse spatial arrays. We mitigate this through cache-aligned Array-of-Structures (AoS) data packing which is only unpacked a single time into pre-allocated pinned memory.

The base terrain map consists of \texttt{alignas(8) Tile} structures, where independent fields such as boolean properties (walkable, flyable), tile types, and agent line-of-sight masks are densely bit-packed into a single \texttt{uint32\_t} field. This way, the CPU can load a tile's complete spatial and semantic profile natively into a single 64-byte L1 cache line alongside its neighboring tiles. 

\begin{table}[h]
\centering
\caption{Physical Memory Layout of a Single Environment Stride}
\begin{tabular}{llll}
\hline
\textbf{Component} & \textbf{Type} & \textbf{Offset Calculation} & \textbf{Size (Bytes)} \\ \hline
Grid & \texttt{Tile[]} & $0$ & $W \times H \times 8$ \\
Agents & \texttt{AgentState[]} & \texttt{offset\_agents} & $N_{agents} \times 24$ \\
POIs & \texttt{POIState[]} & \texttt{offset\_pois} & $N_{pois} \times 16$ \\
Speeds & \texttt{float[]} & \texttt{offset\_speeds} & $N_{agents} \times N_{types} \times 4$ \\
Knowledge & \texttt{Belief[]} & \texttt{offset\_agent\_know} & Variable (Mode Dependent) \\
Counters & \texttt{int[4]} & \texttt{offset\_counters} & 16 \\ \hline
Padding & \texttt{uint8[]} & \texttt{env\_stride} & Aligns to 256-byte boundary \\ \hline
\end{tabular} \label{fig:internal_mem_layout}
\end{table}

\begin{table}[h]
\centering
\caption{Observation Tensor Stride Mapping}
\begin{tabular}{lll}
\hline
\textbf{Tensor Layer} & \textbf{Internal Pointer} & \textbf{Dimension / Logic} \\ \hline
Tile Types & \texttt{TYLE\_TYPE\_START} & $N_{types} \times W \times H$ \\
Altitude & \texttt{ALTITUDE\_TYPE\_START} & $1 \times W \times H$ \\
Detected POIs & \texttt{PIO\_START} & $1 \times W \times H$ (Binary Mask) \\
Discovery Map & \texttt{OBSERVED\_START} & $1 \times W \times H$ (Per-Agent or Global) \\
Self Location & \texttt{MY\_LOCATION\_START} & $1 \times W \times H$ (One-hot) \\
Other Agents & \texttt{OTHER\_LOCATIONS\_START} & $(N-1) \times W \times H$ (Layer-per-agent or 1) \\ \hline
\end{tabular} \label{fig:tansor_mem_layout}
\end{table}

Parallel execution of 128+ environments introduces the substantial hazard of false sharing (introduced in Section~\ref{sec:introduction}). The \texttt{EnvironmentArena} system sidesteps this by calculating the exact contiguous byte-width of an individual environment state (\texttt{raw\_stride}) and forcibly padding it to the nearest 256-byte hardware boundary to defeat aggressive L2 prefetchers on architectures like AMD EPYC~\cite{amdepyc7003hpctg}.

\begin{lstlisting}
env_stride = (raw_stride + 255) & ~255;
\end{lstlisting}

Consequently, Thread 0 computing Environment 0 is strictly hardware-isolated from Thread 1 computing Environment 1. We further optimize NUMA node alignment natively by initializing this pinned memory as \texttt{torch.empty()} buffers in Python, then writing to memory directly inside an OpenMP parallel loop to exploit OS first-touch memory policies~\cite{gureya2020bandwidth} for multi-channel allocation. Finally, all output vectors (such as rewards, termination flags, and truncation signals) buffer intermediate parallel writes onto isolated cache-aligned scratch arrays before sequentially writing to Python-facing contiguous dense memory arrays. This substantially reduces false-sharing invalidation events across CPU interconnect fabrics, enabling near-lock-free vertical scaling across available CPU cores.

Furthermore, we accelerate Python standard randomization loops by directly vectorizing inputs onto the exact target dtypes via \texttt{numpy.random.default\_rng()}, reducing float-casting overhead from the standard \texttt{numpy.random.uniform(-1,1)} call. Even still, random action generation is roughly one-third to half of the total runtime across configurations.

\subsection{Sub-Microsecond Resets via Pristine Grid Templating}
In episodic MARL, environments must be reset continuously as horizons are reached. This engine utilizes a templated \texttt{pristine\_grid} archetype. When an environment terminates, the engine bypasses logical branching and memory allocation during reset by keeping a dense copy of the static parts of the initial state. Utilizing highly optimized contiguous \texttt{memcpy} operations (or vectorized \texttt{std::copy}), the engine simply overwrites the corrupted physical memory of the localized environment with the immutable \texttt{pristine\_grid} copy bytes to substantially reduce wall-clock reset latency, keeping the deep learning framework's DataLoader saturated.

\subsection{Zero-Copy Tensor Architecture}
The common NumPy observation to CUDA memory transfer is expensive. A standard array must be copied into a staging area of pinned memory in RAM before being transferred to VRAM for use on a GPU. To prevent the expensive serialization typical in Python environments, we implement a direct, zero-copy tensor bridge leveraging PyBind11~\cite{pybind11} and PyTorch~\cite{pytorch} pinned memory. PyTorch is used to allocate empty tensors and pass their logical addresses to C++, where individual threads use \texttt{memset} to lazily assign physical address locations via first-touch OS policies by setting logically disjoint memory segments to zero on separate threads, instead of using \texttt{torch.zero()} which would assign all memory to the same controller from the calling thread. This way, the physical memory locations spread across memory controllers for maximum memory bandwidth. This modification benefits systems with highly parallel access patterns while leaving behavior on single-controller systems unchanged.

Once the memory is physically assigned and pinned, the C++ code edits in place before a copy is made to VRAM. When the CPU thread yields back to Python, no arrays are copied, serialized, or instantiated. PyTorch simply interprets its internally mutable buffers and pushes them to the GPU via Direct Memory Access (DMA). This architecture is designed to minimize memory-bus overhead for state transfer to the GPU. Additionally, PyTorch can be limited to one thread, because the GPU does the actual transfer. This way, torch threads and HASE OMP threads do not compete for CPU time. Because a copy occurs to transfer memory to the GPU, the persistent in-place memory is not in danger of being accidentally mutated by operations downstream. If CUDA is not available, the copy to a memory buffer still provides safety from mutation.

\begin{figure}[htbp]
    \centering
    \includegraphics[width=\linewidth]{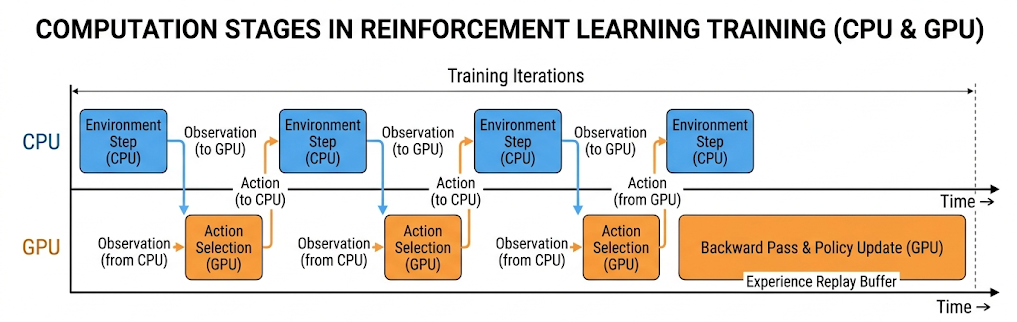}
    \caption{Diagram standard RL pipeline: Env Takes Step(s) (blue), then transfers observation to GPU (blue arrow) to take action(s) (orange)  before those actions are transported back to the CPU (orange arrow). By using parallel environments we amortize each step's overhead cost and by using pinned memory we minimize transfer latency.}
    \label{fig:actionpipeline}
\end{figure}

\subsection{Empirical Optimization Constraints and Engine Trade-Offs}
While standard C++ architectural best practices dictate certain memory and execution paradigms, empirical profiling revealed that several conventional optimizations actively degraded the engine's throughput. Consequently, the following implementation details were deliberately structured to subvert classical advice in favor of measured performance:

\begin{itemize}
    \item \textbf{Native Array Representation over \texttt{std::vector<bool>}:} The C++ standard library specializes \texttt{std::vector<bool>} to pack booleans as proxy bits to save memory. The memory allocation and conversion logic required to pass these arrays back to Python as an array of bools turned out to be slower than the packing was worth. The engine explicitly substitutes all internal masking instances with \texttt{std::vector<uint8\_t>}, to keep memory copy-free.
    \item \textbf{Scattered Tensor Writes over PyTorch Striding:} PyTorch allocates continuous memory blocks linearly based on dimensions \texttt{[envs, channels, height, width]}. Classical Loop Nest Optimization (LNO) dictates that the innermost C++ loop should iterate over the contiguous spatial dimension (\texttt{width}) to march linearly through CPU cache lines for each channel. However, natively conforming to this layout required looping sequentially through $N=$\texttt{n\_tiles} (channels) externally and checking the spatial map internally for each channel iteratively. The cost of $O(NWH)$ complexity of re-scanning the map outweighed worse cache locality at $O(WH)$ complexity.
    \item \textbf{Branch Predictor Reliance over Branchless Execution:} To populate the observation tensors, the C++ backend masks unseen tiles. A common algorithmic optimization is branchless execution casting a boolean visibility check to a float (0.0 or 1.0) and multiplying it against the spatial state to avoid conditional \texttt{if} statements. Profiling demonstrated this was measurably slower. Because visibility states are overwhelmingly \texttt{false} early in an episode and strictly monotonic (once seen, always seen), modern CPU branch predictors efficiently handle the conditional block due to the monotonic nature of the visibility states. Forcing branchless assignment resulted in unneeded, redundant floating-point writes to memory, consuming critical memory bandwidth.
\end{itemize}

\section{Architectural Ablation Results}
\label{sec:ablation}
To quantify the systemic impact of the Data-Oriented Design and zero-copy paradigms on raw environment throughput, we conducted an incremental ablation spanning five critical structural variants.

\begin{figure}
    \centering
    \includegraphics[width=0.5\linewidth]{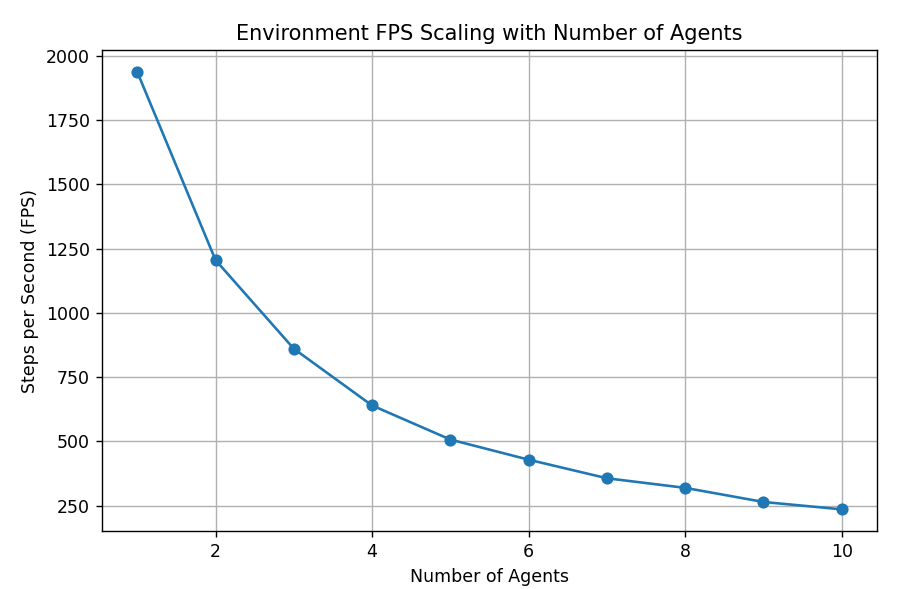}
    \caption{Laptop Speedtest on native single threaded numpy environment}
    \label{fig:numpy}
\end{figure}

\begin{figure}[htbp]
    \centering
    \begin{tabular}{cc}
        \includegraphics[width=0.48\linewidth]{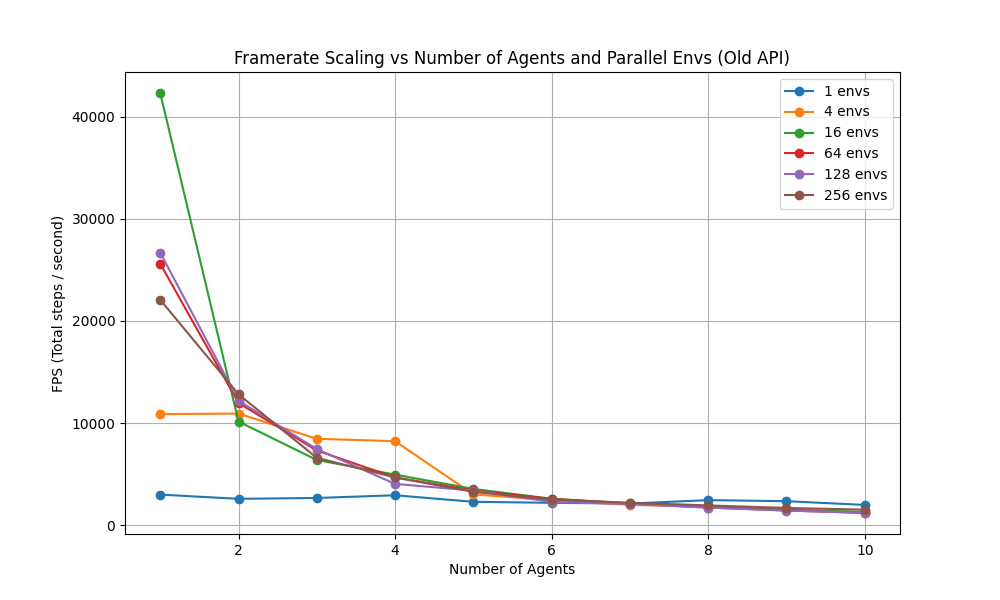} &
        \includegraphics[width=0.48\linewidth]{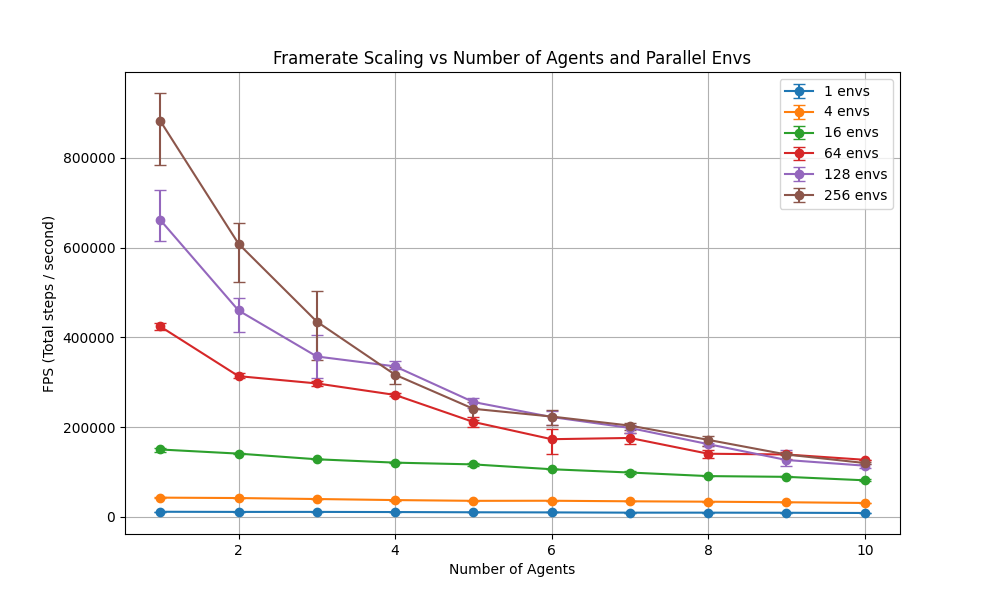} \\
        (a) Baseline (Tensor State) & (b) Zero-copy Cache Aligned \\[6pt]
        \includegraphics[width=0.48\linewidth]{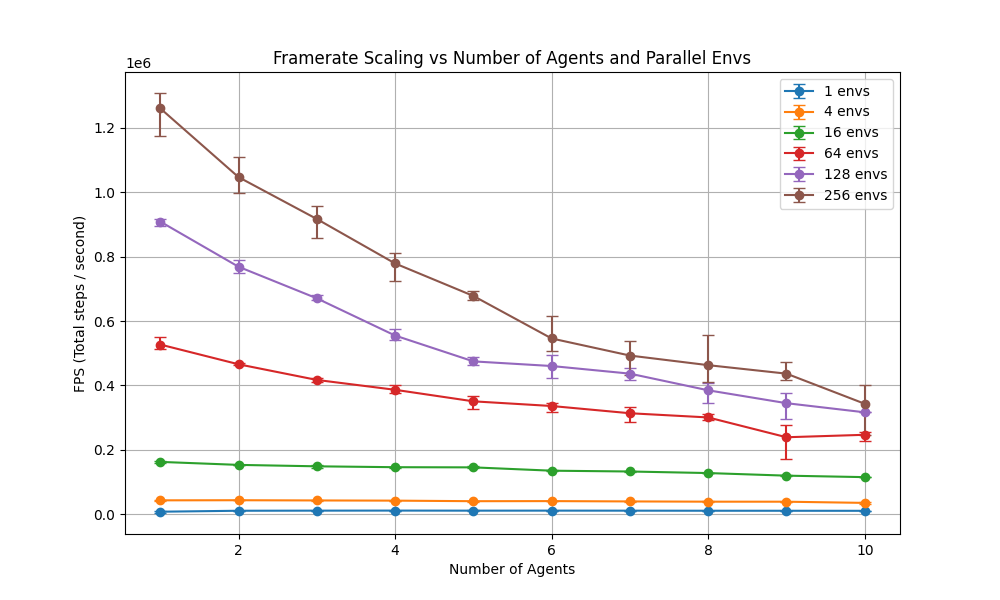} &
        \includegraphics[width=0.48\linewidth]{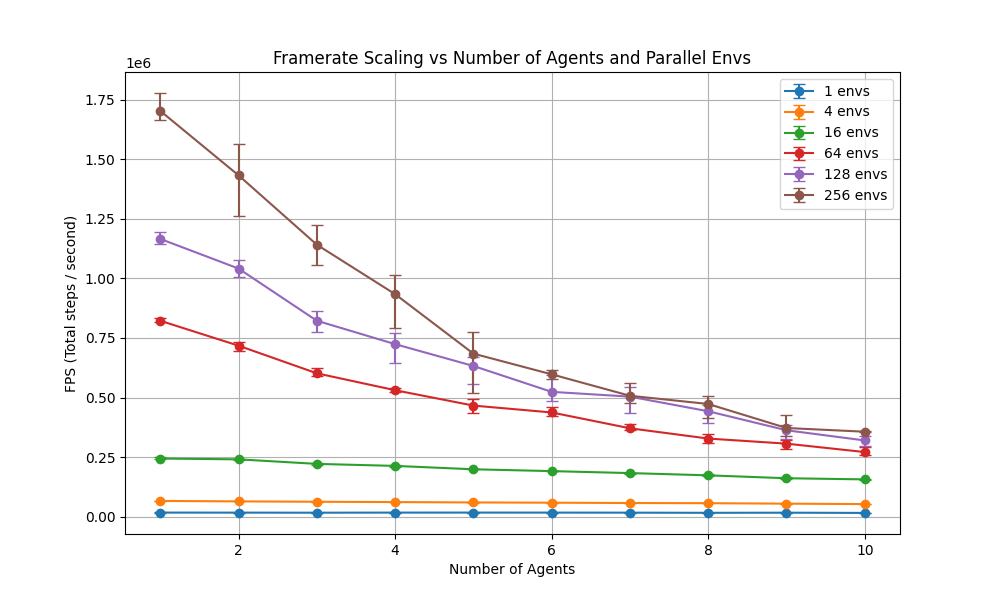} \\
        (c) Diff Sweep & (d) Static Shared Arrays \\[6pt]
        \includegraphics[width=0.48\linewidth]{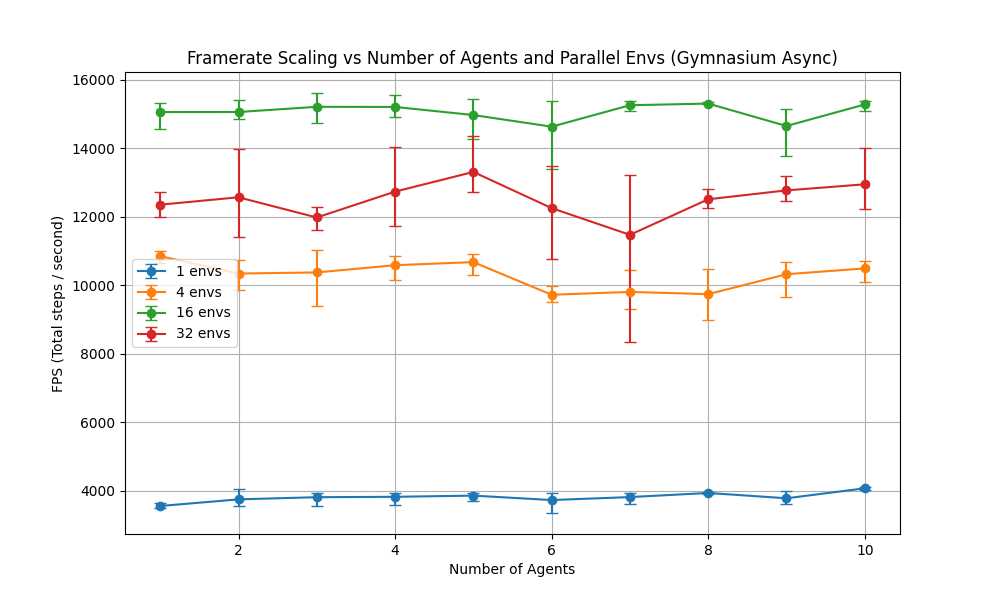} &
        \includegraphics[width=0.48\linewidth]{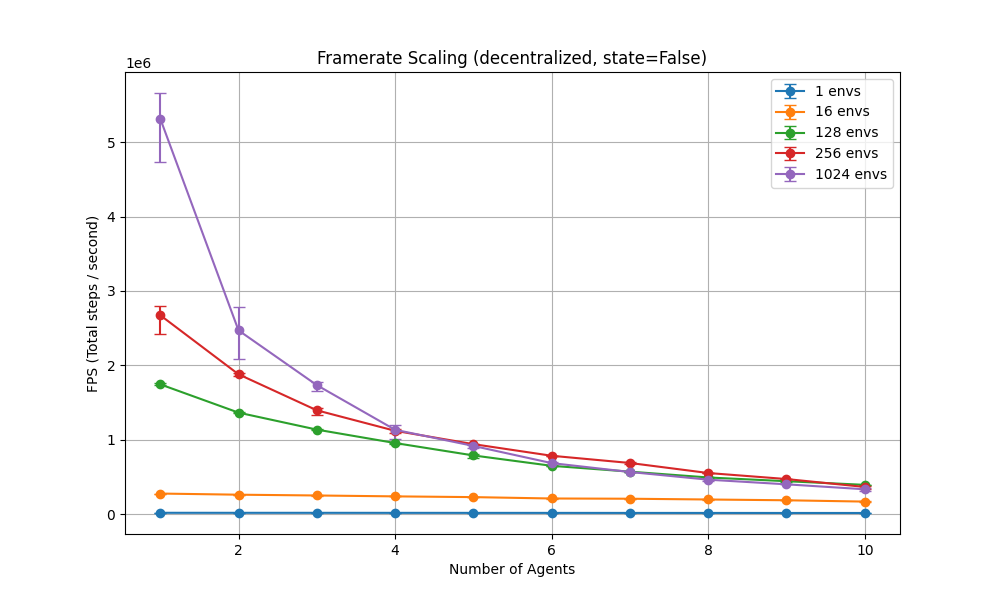} \\
        (e) Gymnasium Async & (f) NUMA \& False-Sharing Local
    \end{tabular}
    \caption{Performance scaling evaluation of progressive architectural improvements against varying environment counts during the ablation study.}
    \label{fig:ablation_study}
\end{figure}

\begin{enumerate}
    \item \textbf{Baseline (Tensor State):} A standard OpenMP-threaded C++ backend maintaining internal environment state as a contiguous float tensor of disjoint single-precision floats manually converted from a \texttt{py::tuple} via \texttt{torch.to()} every step.

    \item \textbf{Zero-copy Cache Aligned:} Internal state is compactly compressed into a locally dense \texttt{uint8} Array of Structures. A unified multi-threaded loop transcribes the full state directly into a pinned PyTorch tensor mapped at initialization.

    \item \textbf{Diff Sweep:} Bypasses explicit dense transcription by computing logical diffs at execution time. Only procedurally updated tiles (e.g., dynamically traversed fog-of-war barriers) issue writes onto the shared tensor array space. In the worst case (fully revealed map), this degenerates to $O(WH)$ writes per step.

    \item \textbf{Static Shared Arrays:} Removes tuple-packing loops for episodic returns. Individual scalars for rewards, and booleans for termination and truncation, are inscribed directly onto shared, statically allocated GPU-visible registers.

    \item \textbf{Gymnasium Async:} Benchmarks against the Python community standard by reverting to purely asynchronous execution via native Gymnasium~\cite{towers2024gymnasium} Vector environments. Typical reinforcement algorithms constrain purely asynchronous engines via severe out-of-memory errors above $M=32$ simultaneous instances.

    \item \textbf{NUMA \& False-Sharing Localization:} Addresses performance bottlenecks on high-core-count processors (e.g., AMD EPYC). We incorporated 256-byte cache-alignment padding alongside thread-local intermediate buffer assignments to substantially reduce L1/L2 cache invalidation events when computing output rewards. The integration of \texttt{np.random.default\_rng()} directly generated action dtypes without recast overhead, achieving peak single-agent throughput of approximately 5,500,000 SPS on a 4-core mobile laptop and approximately 12,800,000 SPS on a 16-core server (see Table~\ref{tab:hardware_comparison}). Scaling to 1024 simultaneous environments yielded substantially higher throughput across all hardware configurations.

    \item \textbf{First-Touch Memory Allocation:} The impact of parallel first-touch initialization versus serial OS memory mapping is quantified in Phase~1 (Section~\ref{sec:phase1}); serial initialization exhibited severe memory bandwidth bottlenecks on the EPYC hardware due to NUMA node imbalance. Also made individual scalars for rewards, and booleans for termination and truncation cache local C++ vectors before a final single-threaded copy to the static tensors to reduce false sharing during rewards accumulation. 
\end{enumerate}

\section{Hardware Optimization and MARL Tuning}
\label{sec:hardware_tuning}

While the DOD architecture reduces intra-environment cache contention, deploying the engine across multi-socket, high-core-count systems introduces additional hardware-level bottlenecks that require empirical tuning. Simulators parallelized via OpenMP~\cite{openmp08} can experience severe performance degradation when orchestrated by Python-based deep learning frameworks (i.e., PyTorch~\cite{pytorch}) on high-core-count microarchitectures (e.g., 2x AMD EPYC 7282 16-Core). Server hardware requires fundamentally different treatment from consumer components due to factors like Non-Uniform Memory Access (NUMA)~\cite{gureya2020bandwidth} discrepancies, where cross-socket lookups incur heavy latency, and thread thrashing when CPU oversubscription disrupts the pipeline interconnect fabrics. To systematically resolve these hardware-level bottlenecks, we conducted a comprehensive, two-phase optimization study.

\subsection{Phase 1: Pure C++ MARL Environment Benchmarks}
\label{sec:phase1}
Before analyzing PyTorch overhead, it was critical to establish a baseline for the localized C++ engine's parallel execution utilizing a centralized state configuration.

\subsubsection{Hypotheses and Experiments}
The primary hypothesis for this phase was that hardware-enforced thread pinning alongside first-touch memory allocation (initializing memory by the thread that subsequently operates on it) would maximize performance by forcing the physical memory pages onto the local memory controller and avoiding cross-socket NUMA lookups.

To effectively examine this hypothesis, we evaluated three configurations spanning 1,000,000 algorithmic steps:
\begin{enumerate}
    \item \textbf{Control (OS-Managed):} Standard OS-managed CPU scheduling allowing the system to fluidly balance OpenMP threads by a thread pool where each environment step is assigned to any currently free thread.
    \item \textbf{Thread Pinning:} Strict bindings explicitly mapping OpenMP threads to distinct physical cores utilizing \texttt{OMP\_PLACES=cores OMP\_PROC\_BIND=true}.
    \item \textbf{Serial Init:} Memory initialized on a single thread to strictly avoid any parallel first-touch placements.
\end{enumerate}

\subsubsection{Pinning Results}
\begin{figure}[htbp]
    \centering
    \begin{tabular}{cc}
        \includegraphics[width=0.48\linewidth]{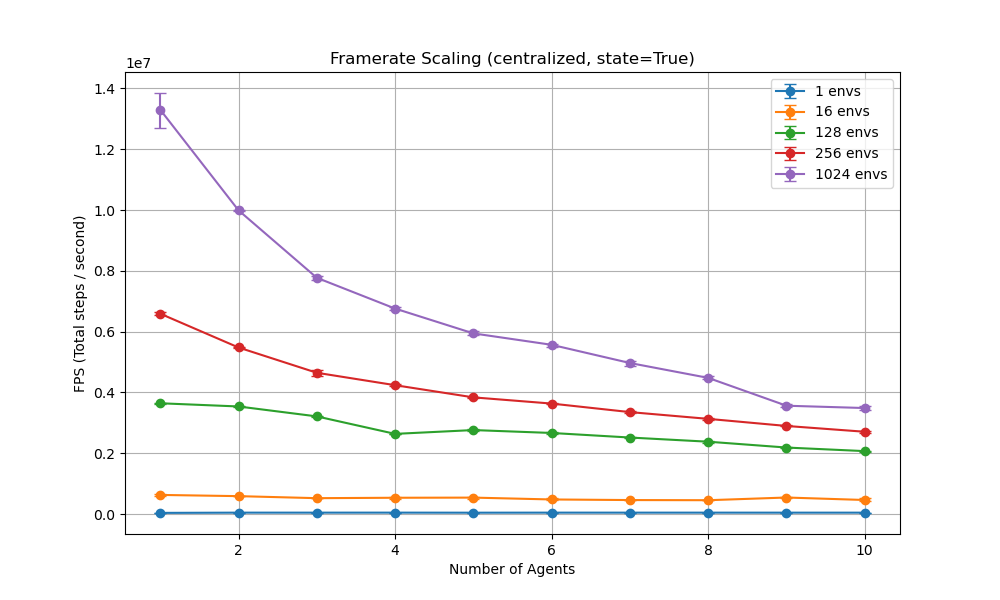} &
        \includegraphics[width=0.48\linewidth]{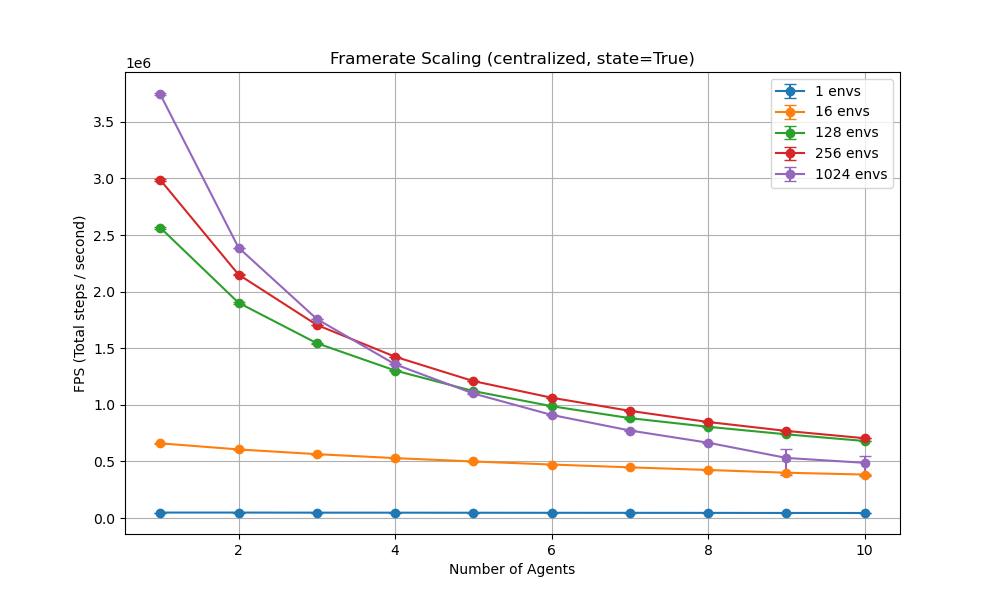} \\
        (a) OS-Managed Control & (b) Strict Thread Pinning \\[6pt]
        \multicolumn{2}{c}{\includegraphics[width=0.48\linewidth]{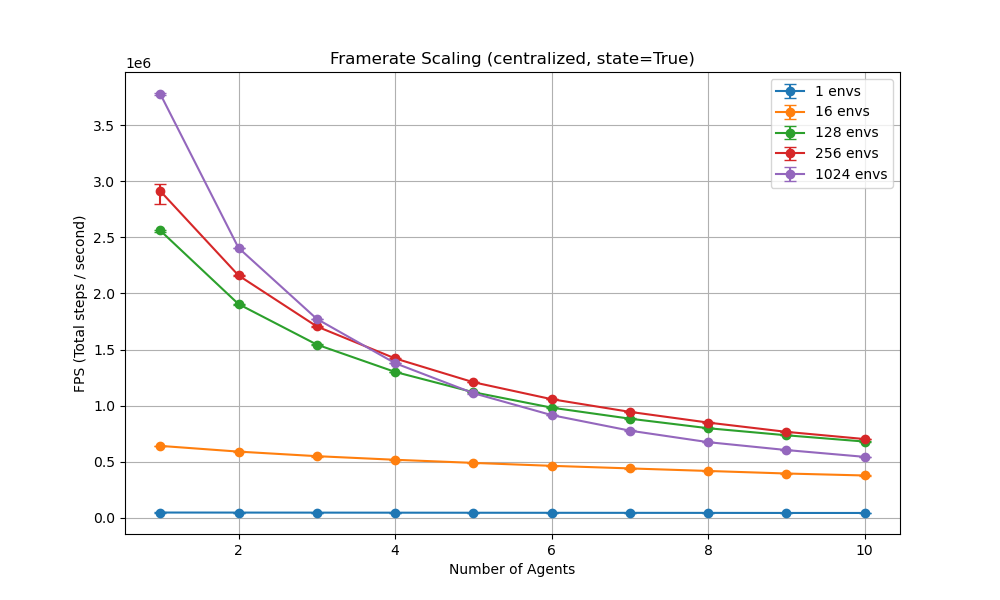}} \\
        \multicolumn{2}{c}{(c) Serial Initialization (No First-Touch)} \\
    \end{tabular}
    \caption{Phase 1 throughput comparison for C++ execution utilizing different core-affinity and initialization strategies on AMD EPYC. Strict thread pinning severely reduced total SPS throughput.}
    \label{fig:phase1_results}
\end{figure}

The hypothesis was not fully supported. The \textbf{Control (OS-Managed)} completed the benchmarks between 2$\times$ and 4$\times$ faster than the strictly pinned equivalent (Figure~\ref{fig:phase1_results}). This result is consistent with the fact that strictly pinning OpenMP to physical cores (\texttt{OMP\_PLACES=cores}) limits execution to only the 32 physical cores available on the system. The OS-managed control is free to leverage all 64 logical threads (including Simultaneous Multithreading), naturally leading to higher total CPU throughput without inducing unnecessary stalls. Additionally, not all environment steps take the same amount of time; for example, radio communication incurs additional computational cost so that static allocation wastes time when quicker threads must wait for slower ones instead of picking up an available job when functioning as a free thread pool.

However, the necessity of first-touch memory allocation was supported by the data. The \textbf{Serial Init} strategy exhibited severe performance degradation due to Infinity Fabric congestion. Because it was initialized serially, the memory arrays failed to distribute evenly across multiple NUMA nodes. Consequently, when all active threads later requested parallel evaluation, they saturated a single memory controller interconnect, cascading into a severe memory bandwidth bottleneck.

\subsection{Phase 2: PyTorch and C++ Integration}

\subsubsection{Hypotheses and Experiments}
Combining a highly parallelized C++ pipeline with PyTorch introduces thread thrashing. PyTorch natively spawns inter-op threads for tensor operations that contend with the C++ OpenMP simulator threads for CPU time. We hypothesized that strictly restricting PyTorch threading while preventing OpenMP's default active spin-waiting would resolve context-switch thrashing and cleanly yield execution back to GPU scheduling threads. By using pinned memory for DMA, the pytorch CPU thread is not responsible for performing any large memory transfers, but rather it just has to assign those transfers to the GPU. For this reason, pytorch does not need many threads.

A disciplined methodology was established for phase two benchmarking:
\begin{enumerate}
    \item Standardized simulations normalized to 1,000,000 algorithmic steps.
    \item Identical execution sequences repeated six times across each test configuration.
    \item Initial 5-second GPU utilization periods explicitly dropped to filter out memory allocation and initialization artifacts.
\end{enumerate}

The test matrix incorporated the following configurations:
\begin{itemize}
    \item \textbf{Test A (Thread Limit):} Restricting PyTorch inter-op threads (\texttt{torch.set\_num\_threads(1)}).
    \item \textbf{Test B (Yield Cores):} Modifying OpenMP passive states (\texttt{OMP\_WAIT\_POLICY=passive}).
    \item \textbf{Test C (Limit + Yield):} Imposing both passive OpenMP threading and PyTorch thread limits simultaneously.
\end{itemize}

\subsubsection{RL Results}
The hypothesis regarding thread yield management was supported by the data. Setting \sloppy{ \texttt{OMP\_WAIT\_POLICY=passive} } explicitly arrests OpenMP threads from busy-waiting when temporarily idle. Combining this with restricted PyTorch threading (Test C) lowered context-switch thrashing and elevated absolute SPS bounds when utilizing all available cores.

Additionally, targeted structural mapping utilizing \texttt{taskset} locking simulation jobs to the localized CPU PCIe-complex of the assigned GPU was tested. While the ``BEST'' (Aligned + Test C) scenario achieved 18,331 SPS compared to the unpinned Test C peak of 18,904 SPS, this constrained setup only utilized 32 threads (half the logical hardware processors). This demonstrates that while unconstrained execution on all available cores yields the highest total systemic throughput, proper hardware pinning yields substantially higher per-thread throughput.

\begin{table}[htbp]
\centering
\resizebox{\textwidth}{!}{
\begin{tabular}{l c c l r r}
\toprule
\textbf{Configuration} & \textbf{OMP Wait} & \textbf{Threads} & \textbf{Affinity} & \textbf{SPS ($m \pm SEM$)} & \textbf{GPU Util ($m \pm SEM$)} \\
\midrule
\multicolumn{6}{l}{\textit{Dual NUMA Nodes / Full System (64 Threads)}} \\
\midrule
\textbf{Control} (Default settings) & Default & 0 & N/A        & $18,379 \pm 171.9$          & $40.3\% \pm 0.37\%$ \\
\textbf{Test A} (Thread Limit)      & Default & 1 & N/A        & $18,829 \pm 50.2$           & $38.5\% \pm 0.29\%$ \\
\textbf{Test B} (Yield Cores)       & Passive & 0 & N/A        & $17,719 \pm 70.6$           & $37.3\% \pm 0.45\%$ \\
\textbf{Test C} (Limit + Yield)     & Passive & 1 & N/A        & $\mathbf{18,904 \pm 68.6}$  & $38.4\% \pm 0.29\%$ \\
\midrule
\multicolumn{6}{l}{\textit{Single NUMA Node / Constrained (32 Threads)}} \\
\midrule
\textbf{Aligned} (PCIe nodes)       & Default & 1 & Aligned    & $18,889 \pm 66.5$           & $38.5\% \pm 0.33\%$ \\
\textbf{Misaligned} (Cross-Fabric)  & Default & 1 & Misaligned & $18,133 \pm 67.8$           & $39.1\% \pm 0.33\%$ \\
\textbf{BEST} (Aligned + Test C)    & Passive & 1 & Aligned    & $18,331 \pm 69.0$           & $38.2\% \pm 1.06\%$ \\
\textbf{WORST} (Misaligned + Test B)& Passive & 0 & Misaligned & $16,600 \pm 8.2$            & $36.8\% \pm 0.94\%$ \\
\bottomrule
\end{tabular}
}
\caption{Hardware optimization result matrix comparing CPU contention variables against SPS throughput and GPU utilization. Values represent $Mean \pm SEM$ for $n=6$ trials. The 32-thread subset demonstrates higher per-core efficiency by avoiding Cross-Fabric interconnect latency.}
\label{tab:results_matrix}
\end{table}

\subsection{Deployment Recommendations}

For maximizing total absolute throughput on a given machine, avoid thread pinning, use only the GPU aligned cores, enforce parallel first-touch memory allocation, and limit PyTorch thread count (\texttt{torch.set\_num\_threads(1)}). For multi-NUMA, multi-GPU architectures, we recommend strictly mapping simulation jobs to the localized CPU PCIe-complex of the assigned GPU via \texttt{taskset} and then turning OMP passive wait on if all machine cores are in use (starvation possibility) and off if there are other logical cores which torch can use. In practice, this constraint barelychanges raw overall maximum SPS per job while keeping a full CPU free. Maximizing per-thread efficiency ensures multiple concurrent jobs optimally saturate independent device throughput across the platform without overloading interconnects.

\section{Reinforcement Learning Methodology}
\label{sec:methodology}

Having established the engine's raw throughput, we now validate that this throughput translates into practical RL training gains by evaluating three model-free algorithms on the Hide-And-Seek task. To effectively exploit the multi-modal observation spaces inherently generated by POMDP environments, our framework incorporates specialized neural architectures for feature extraction. The core algorithms are Multi-Agent Proximal Policy Optimization (MAPPO)~\cite{yu2022surprising}, Branching-Dueling Q-Networks (BDQ)~\cite{tavakoli2018action}, and Multi-Agent Soft Actor-Critic (MASAC)~\cite{haarnoja2018soft}. All utilize a Mixed CNN-Logical Encoder for the spatial portion of the observation space.

\subsection{Mixed Observation Encoder Architecture}
At each operational step, an agent receives a spatial 3D observation tensor capturing localized line-of-sight terrain semantics alongside a dense 1D vector detailing internal telemetry (e.g., coordinates, status). The mixed encoder explicitly bifurcates these inputs. The spatial tensor is routed through three sequential CNN convolutional layers parameterized with kernel sizes $\langle8, 4, 3\rangle$ and strides $\langle4, 2, 1\rangle$ respectively, followed by a flattening fully-connected linear operation to compress spatial invariants. Concurrently, the logical tensor is projected via an independent Multi-Layer Perceptron (MLP). The resultant feature vectors are concatenated before being fed into either the shared critic (for CTDE) or independent actor-critic models.

\subsection{Reinforcement Learning Algorithms}
To evaluate the convergence properties and computational throughput of the engine, we trained agents using three distinct model-free reinforcement learning algorithms. Each algorithm handles continuous environment steps by translating observed encoded states into discrete multi-agent actions, though they differ substantially in optimization mechanics (e.g., on-policy vs.\ off-policy).

\subsubsection{Proximal Policy Optimization (PPO)}
PPO~\cite{schulman2017proximal} is an on-policy actor-critic algorithm. Our implementation (MAPPO/IPPO\cite{yu2022surprising}) uses an actor that outputs categorical distributions over the discrete navigation and radio action spaces and a shared critic estimating $V(s)$. Rollouts of fixed length (128 steps * num environments) are batched across vectorized environments and consumed to compute Generalized Advantage Estimation (GAE). Policy updates are bounded by a clipping function ($\epsilon=0.2$) to limit KL divergence from the previous policy.

\subsubsection{Deep Q-Network (DQN)}
DQN~\cite{mnih2015human} formulates an off-policy, value-based approach with $\varepsilon$-greedy exploration. Our formulation employs a Branching-Dueling architecture with split advantage heads for movement and radio actions summed against a shared state-value baseline. Transitions $\langle s, a, r, s', d \rangle$ are stored in a centralized replay buffer of capacity $10^6$ transitions. Mini-batches are sampled to minimize temporal difference (TD) error against Bellman targets; a synchronized target network stabilizes value estimation during long search episodes.

\subsubsection{Soft Actor-Critic (SAC)}
SAC~\cite{haarnoja2018soft} bridges off-policy efficiency with entropy regularization. We implement a discrete variant suitable for multi-agent grid navigation, maintaining a replay buffer of capacity $10^6$ transitions. The actor outputs an entropy-regularized policy $\pi(a|s)$ and paired soft Q-networks mitigate positive bias; the temperature $\alpha$ is auto-tuned to balance exploration and exploitation.

\subsection{Learning Validation and Convergence}
\begin{figure}[htbp]
    \centering
    \includegraphics[width=\linewidth]{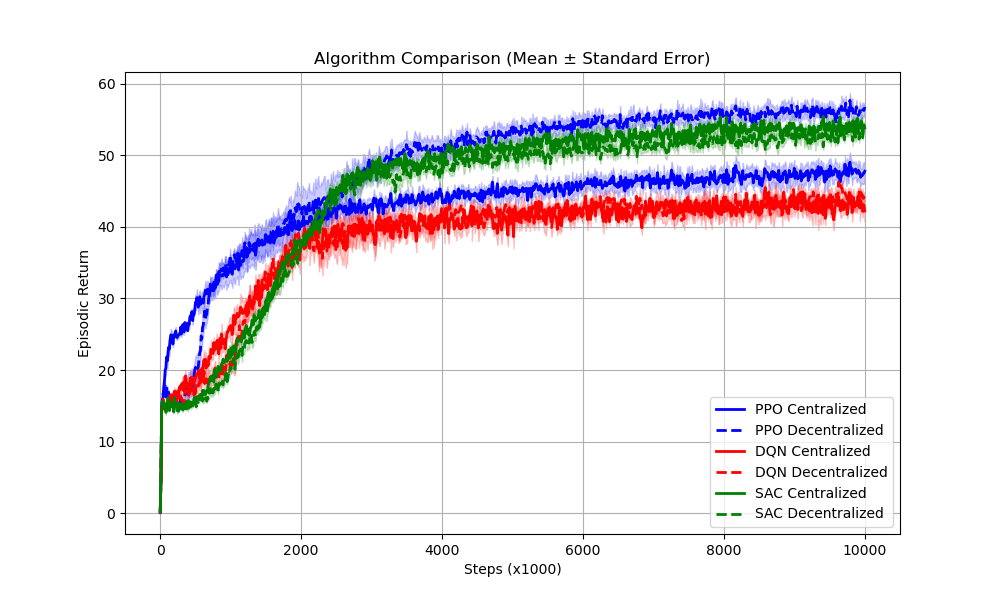}
    \caption{Episodic return trajectories demonstrating the sample complexity demands of various architectures operating over 1 million steps in highly parallelized zero-copy environments.}
    \label{fig:combined_learning_curves}
\end{figure}

Figure~\ref{fig:combined_learning_curves} displays the results of 6 independent training runs across the various RL architectures. Both SAC and DQN demonstrate similar convergence behaviors regardless of whether models operate in decentralized or centralized observation modes, though centralized architectures run approximately $n_\text{agent}$ times slower because each agent's experience must be processed through the network independently. Conversely, PPO scales more favorably when utilizing independent per-agent models (IPPO) compared to a monolithic centralized model (MAPPO). On the 9950X with 128 parallel environments, the environment steps account for approximately 10 seconds of the total 400~s (PPO), 1200~s (DQN), and 2400~s (SAC) training runtimes for $10^7$ frame runs, indicating that environment throughput is no longer a bottleneck for these algorithms.

\section{Conclusion}
\label{sec:conclusion}
We presented Hide-And-Seek-Engine, a compute-efficient Dec-POMDP simulator natively architected in C++ for multi-agent search scenarios. Through Data-Oriented Design, 64-byte cache-line alignment, and a zero-copy PyTorch memory bridge via pinned memory and DMA, the engine achieves throughputs exceeding 33,000,000 SPS in single-agent configurations on an AMD Ryzen 9950X, representing a throughput increase of approximately 3,000$\times$ over a NumPy baseline. A two-phase hardware optimization study identified first-touch NUMA memory allocation and OpenMP thread yield policies as critical for high-core-count server deployments. Learning validation confirmed that PPO, DQN, and SAC can all train cooperative MARL policies within the engine, with environment steps consuming less than 3\% of total training wall-clock time in the tested configurations. Future work includes GPU-native hider policies, extended map-scale benchmarks, and integration with existing MARL libraries.

\newpage % Start appendix on a new page
\appendix
\section{Appendix}

\subsection{Extended Hardware Runtime Comparisons}
\label{app:hardware_runtimes}
This appendix details the runtime performance scaling and throughput metrics across varying hardware architectures.

The engine was benchmarked across three hardware profiles to capture scaling limits from low-end mobile cpu constraints to high-end enterprise servers:
\begin{itemize}
    \item \textbf{High-End Desktop (AMD Ryzen 9950X):} 16 Cores, representing peak consumer single-thread performance and monolithic L3 cache topologies (see \texttt{Epyc\_Tuning/env\_configs\_9950X}).
    \item \textbf{Server Architecture (AMD EPYC 7282):} 16 Cores with an RTX Ada 6000 GPU, representing complex NUMA domains, distributed Infinity Fabric interconnects, and strict PCIe pathways (see \texttt{Epyc\_Tuning/env\_configs\_epyc}).
    \item \textbf{Mobile Platform (4-Core Intel Core i5-8300H):} Representing low-end constrained thermal and power footprint hardware (see \texttt{env\_configs\_laptop}).
\end{itemize}

\begin{figure}[htbp]
    \centering
    \begin{tabular}{cc}
        \includegraphics[width=0.48\linewidth]{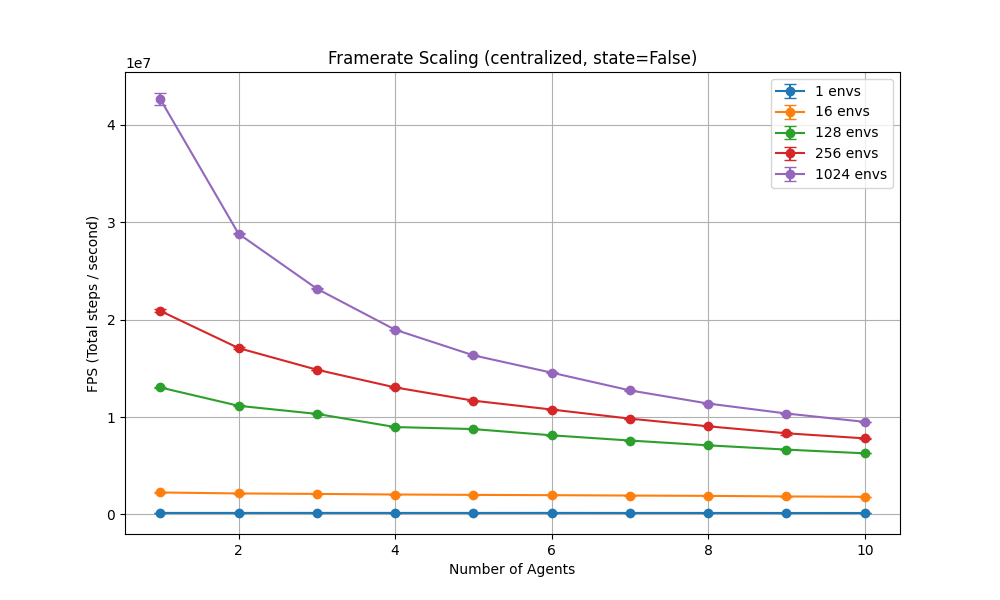} &
        \includegraphics[width=0.48\linewidth]{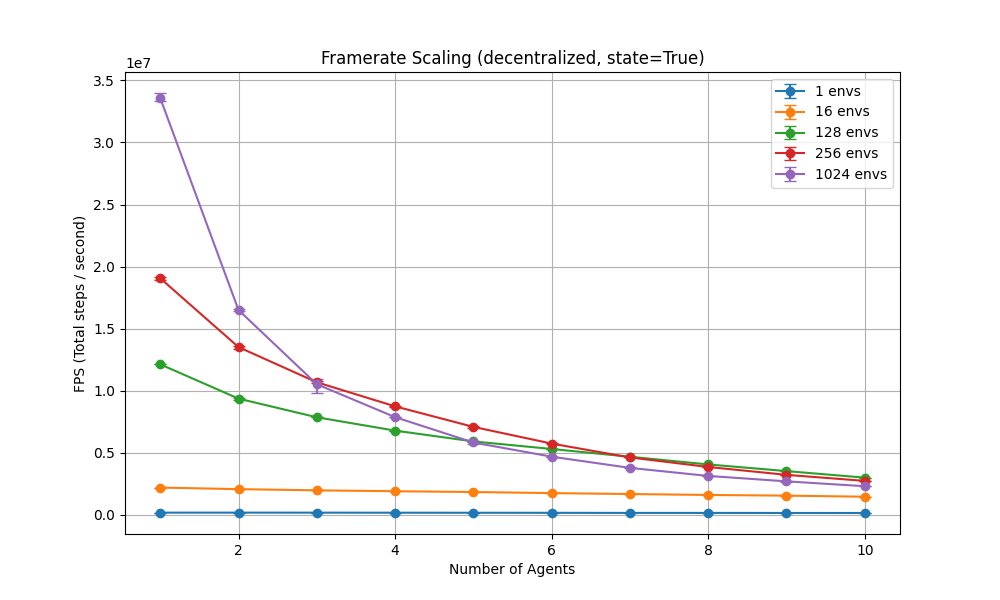} \\
        (a) Centralized Partial Observations & (b) Decentralized Partial Observations \\[6pt]
        \includegraphics[width=0.48\linewidth]{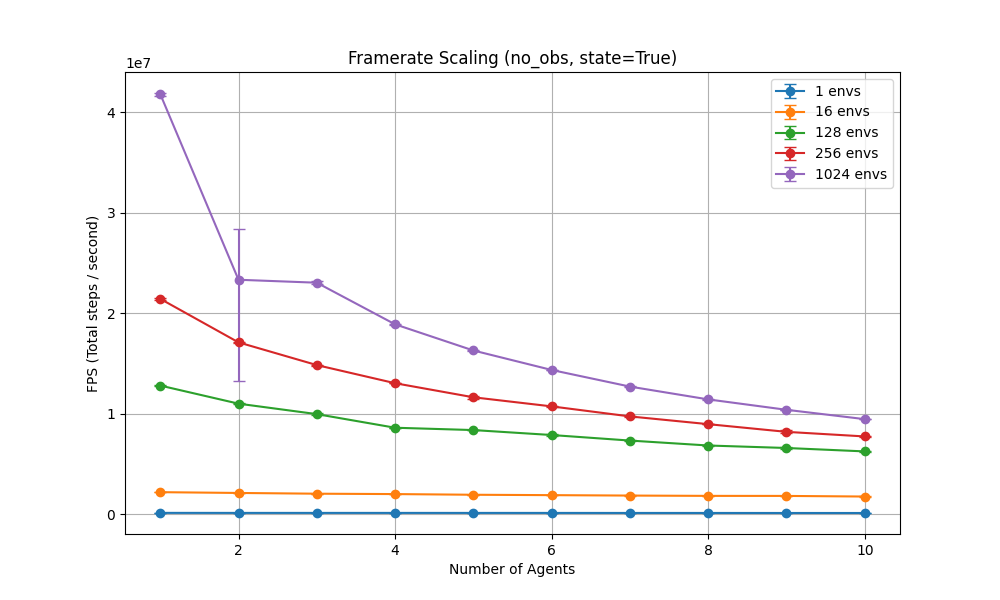} &
        \includegraphics[width=0.48\linewidth]{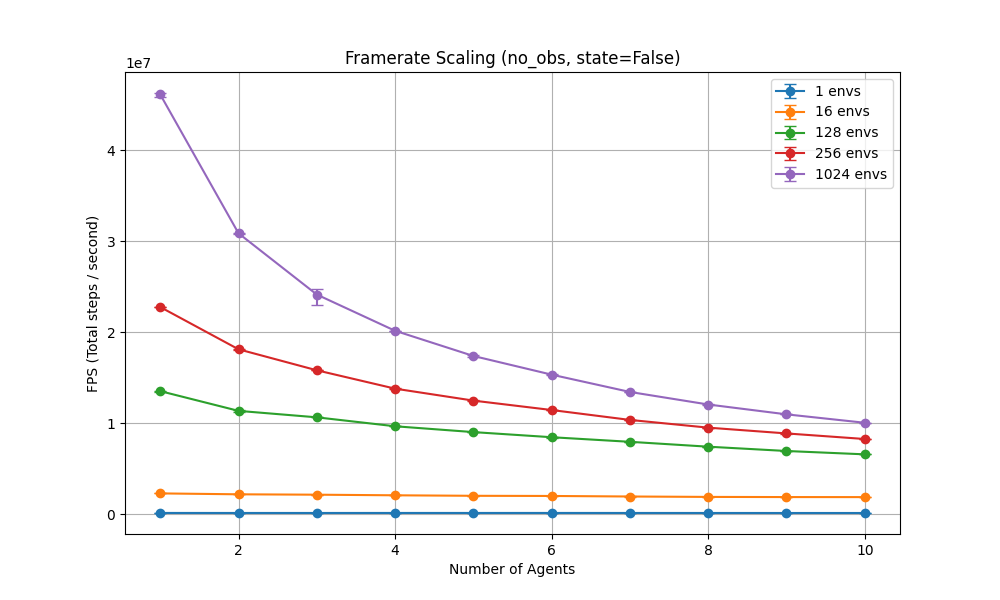} \\
        (c) Fully Observable Centralized State & (d) Internal Engine Stepping (No Tensor Return) \\[6pt]
        \includegraphics[width=0.48\linewidth]{Epyc_Tuning/env_configs_9950X/speedtest_results_centralized_nostate.png} &
        \includegraphics[width=0.48\linewidth]{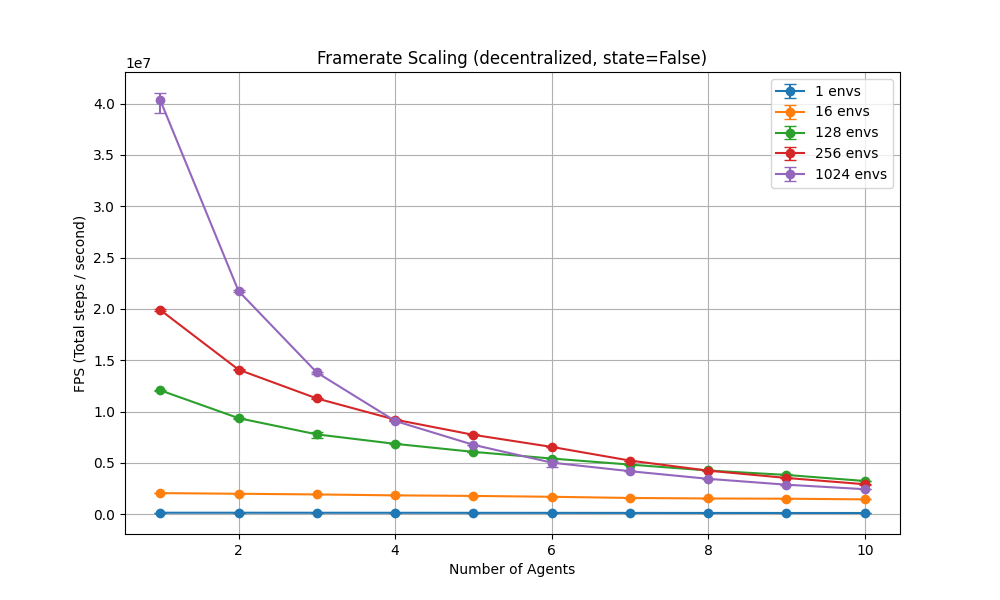} \\
        (e) No Observable Grid Return (Internal Only) & (f) Local Decentralized Logic Data
    \end{tabular}
    \caption{Throughput scaling profiles for the AMD Ryzen 9950x.}
    \label{fig:throughput_scaling_9950x}
\end{figure}

\begin{figure}[htbp]
    \centering
    \begin{tabular}{cc}
        \includegraphics[width=0.48\linewidth]{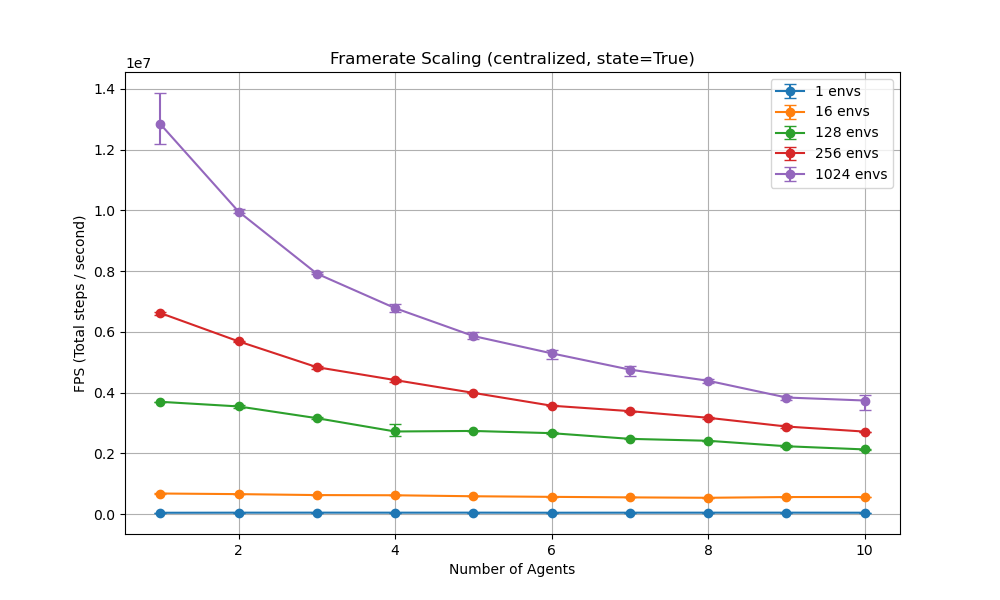} &
        \includegraphics[width=0.48\linewidth]{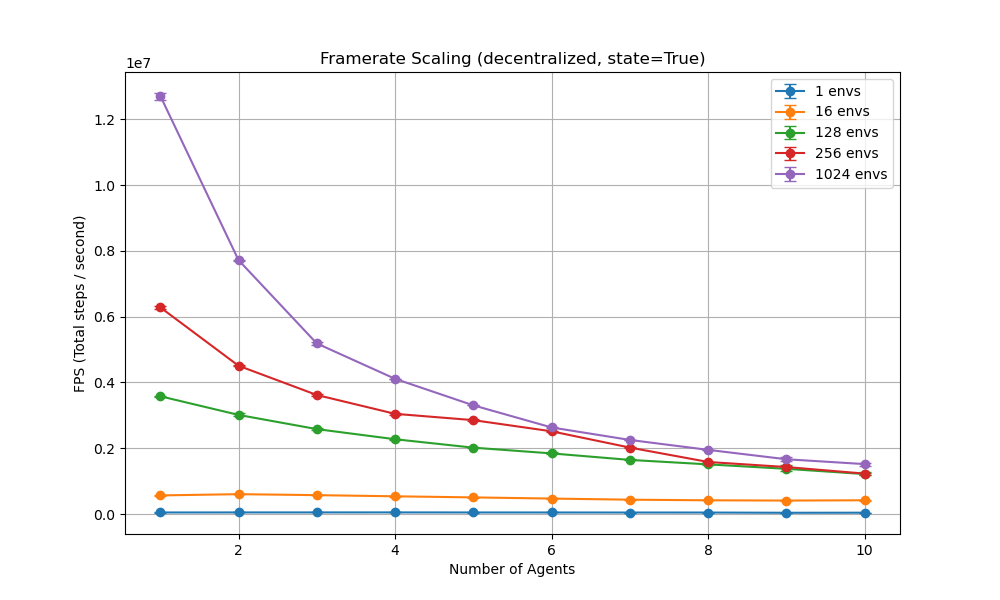} \\
        (a) Centralized Partial Observations & (b) Decentralized Partial Observations \\[6pt]
        \includegraphics[width=0.48\linewidth]{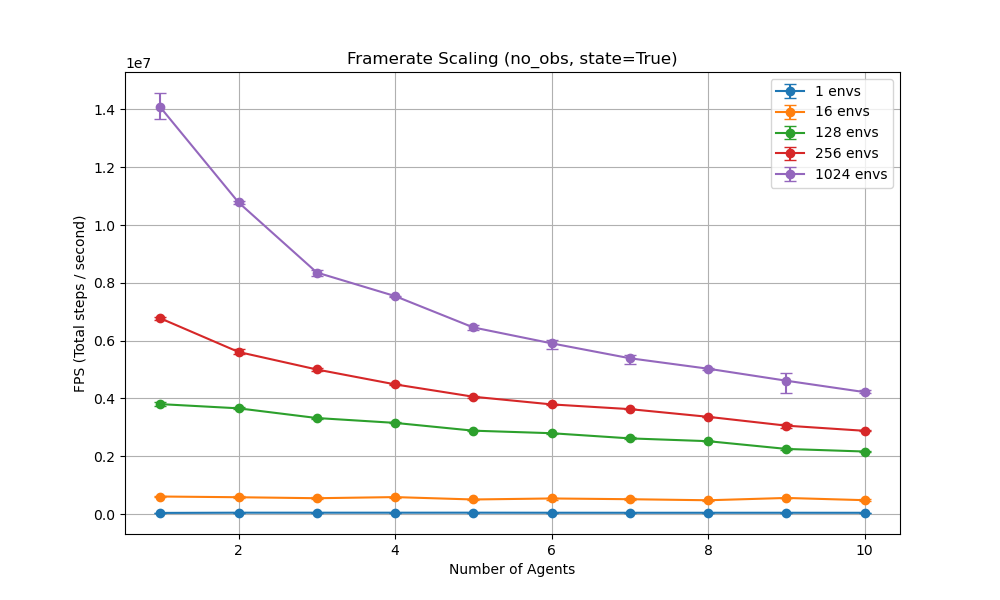} &
        \includegraphics[width=0.48\linewidth]{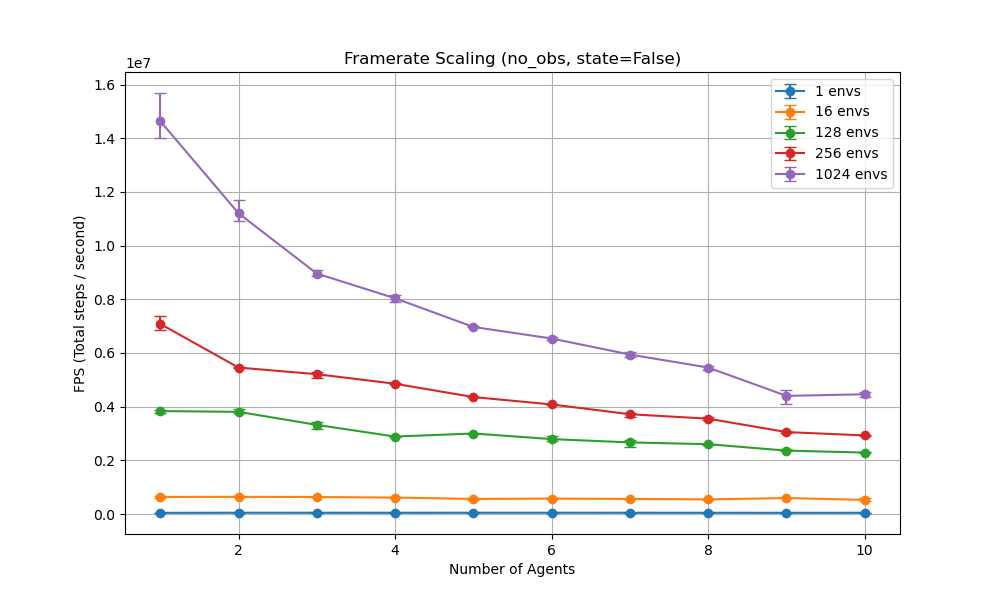} \\
        (c) Fully Observable Centralized State & (d) Internal Engine Stepping (No Tensor Return) \\[6pt]
        \includegraphics[width=0.48\linewidth]{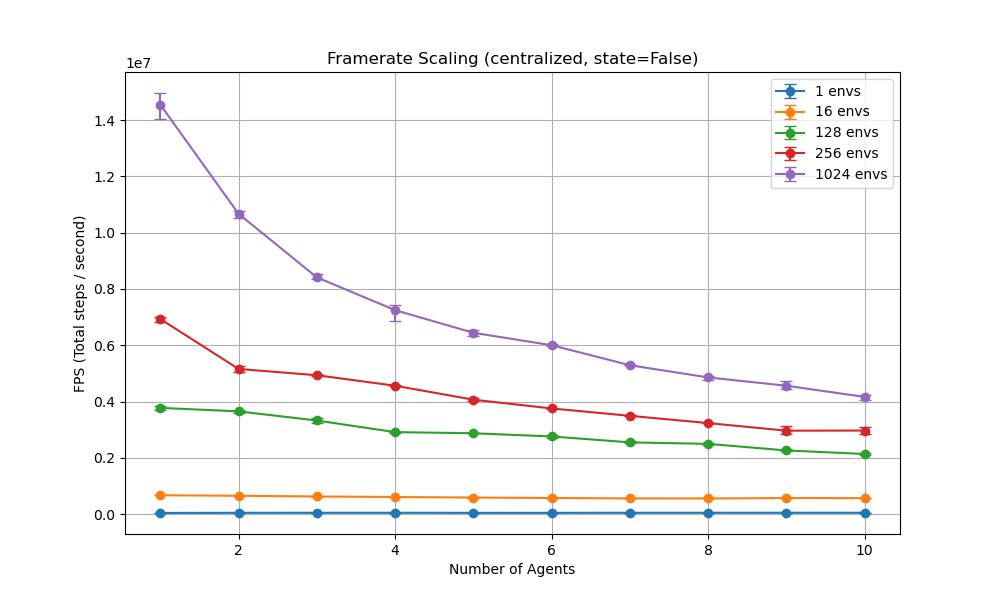} &
        \includegraphics[width=0.48\linewidth]{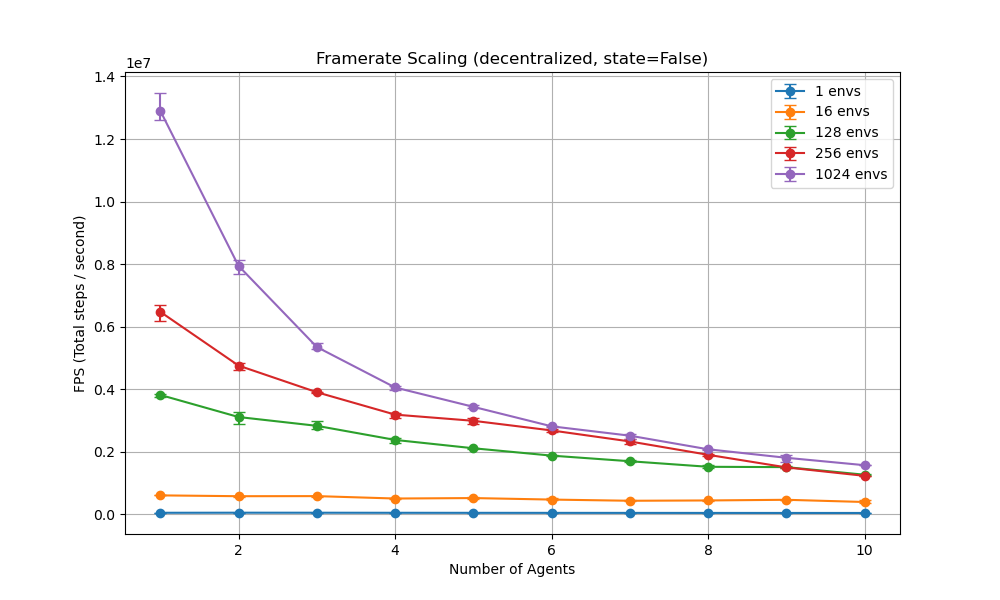} \\
        (e) No Observable Grid Return (Internal Only) & (f) Local Decentralized Logic Data
    \end{tabular}
    \caption{Throughput scaling profiles for the AMD EPYC 7282.}
    \label{fig:throughput_scaling_epyc}
\end{figure}

\begin{figure}[htbp]
    \centering
    \begin{tabular}{cc}
        \includegraphics[width=0.48\linewidth]{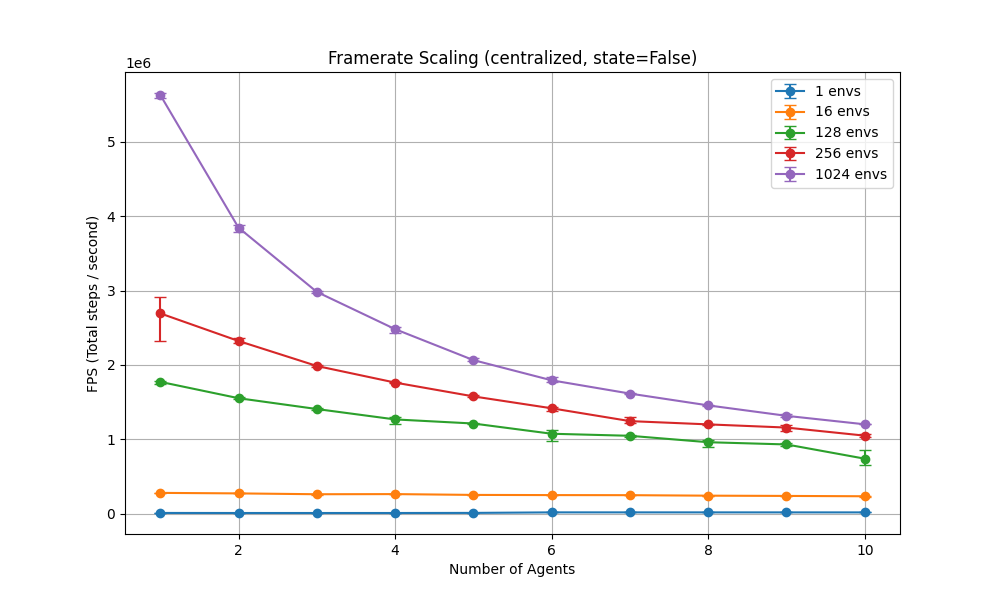} &
        \includegraphics[width=0.48\linewidth]{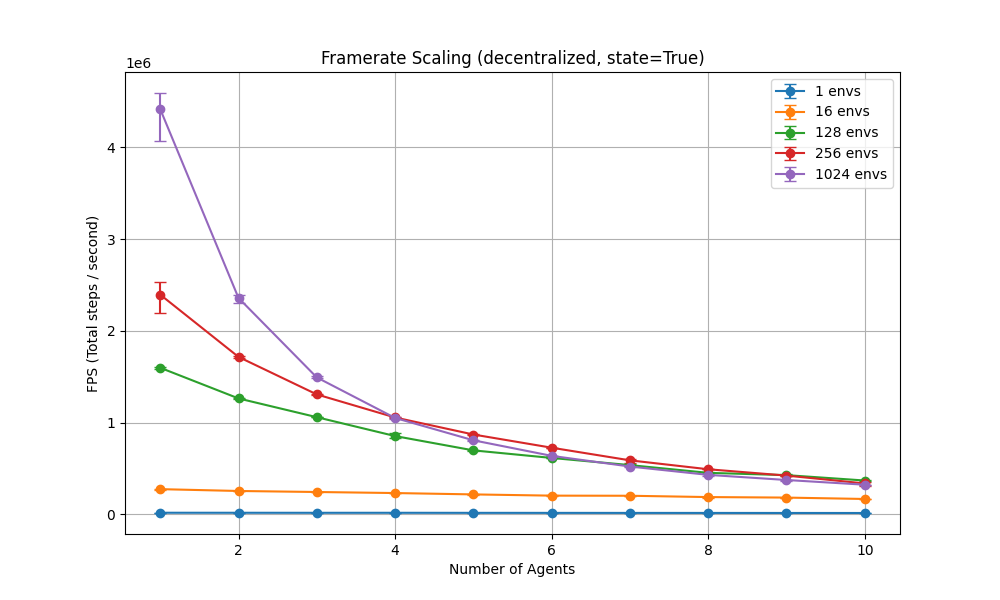} \\
        (a) Centralized Partial Observations & (b) Decentralized Partial Observations \\[6pt]
        \includegraphics[width=0.48\linewidth]{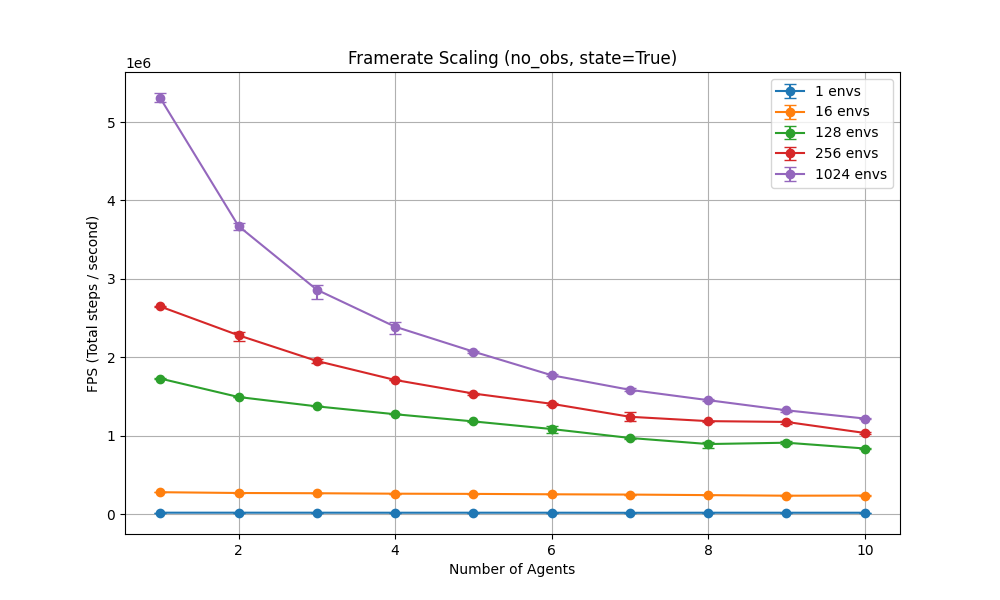} &
        \includegraphics[width=0.48\linewidth]{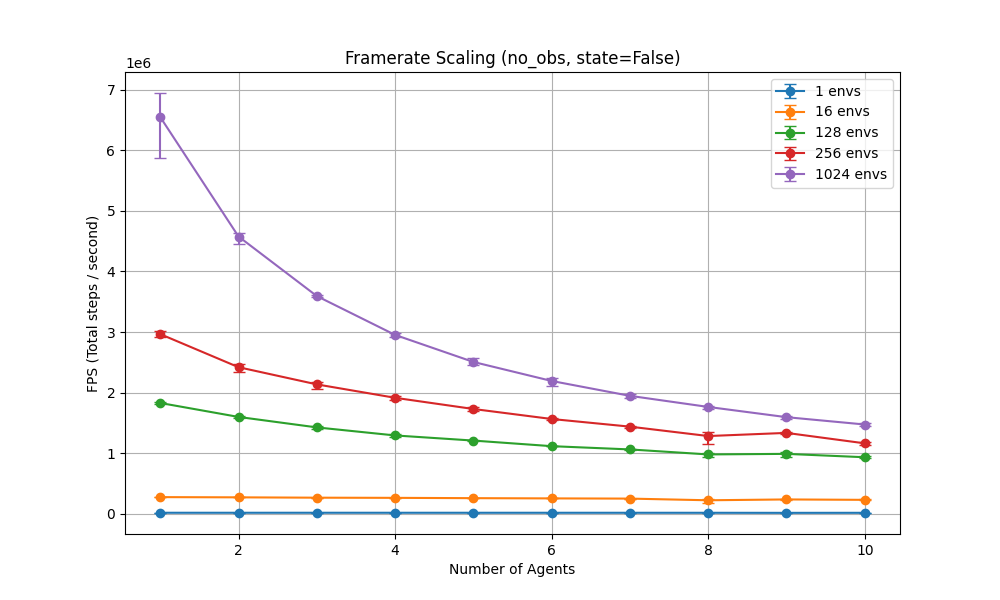} \\
        (c) Fully Observable Centralized State & (d) Internal Engine Stepping (No Tensor Return) \\[6pt]
        \includegraphics[width=0.48\linewidth]{Epyc_Tuning/env_configs_laptop/speedtest_results_centralized_nostate.png} &
        \includegraphics[width=0.48\linewidth]{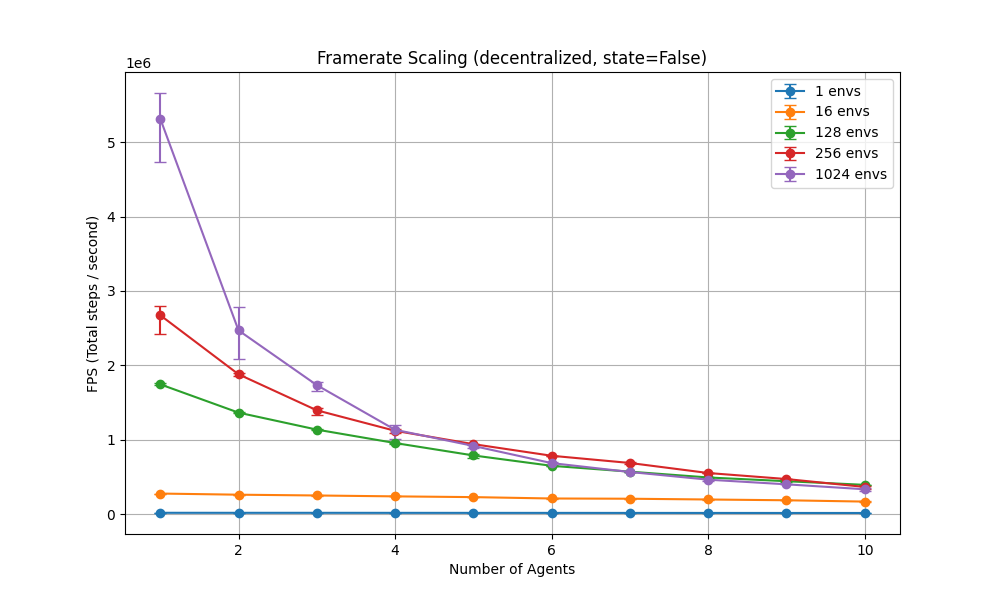} \\
        (e) No Observable Grid Return (Internal Only) & (f) Local Decentralized Logic Data
    \end{tabular}
    \caption{Throughput scaling profiles for the Intel 4-Core Laptop.}
    \label{fig:throughput_scaling_laptop}
\end{figure}

Tables and throughput plots associated with these configurations show throughput scaling across agent counts and observability modes (Figures~\ref{fig:throughput_scaling_9950x}--\ref{fig:throughput_scaling_laptop}).

\begin{table}[htbp]
\centering
\begin{tabular}{lrrr}
\toprule
\textbf{Hardware} & \textbf{1 Agent (SPS)} & \textbf{5 Agents (SPS)} & \textbf{10 Agents (SPS)} \\
\midrule
AMD Ryzen 9950X & 33,986,275 & 7,135,957 & 2,994,134 \\
AMD EPYC 7282 & 12,801,442 & 3,309,042 & 1,564,790 \\
Intel 4-Core Laptop & 5,539,430 & 2,124,718 & 1,236,755 \\
\bottomrule
\end{tabular}
\caption{Peak throughput (SPS) by agent count across hardware configurations (Decentralized, No Observable State mode).}
\label{tab:hardware_comparison}
\end{table}

\subsection{Hyperparameters Configuration}
\label{app:hyperparameters}
Table~\ref{tab:hyperparams} specifies the primary tuning configurations for the three reinforcement learning algorithms utilized during evaluation.

\begin{table}[htbp]
\centering
\begin{tabular}{lccc}
\toprule
\textbf{Parameter} & \textbf{PPO} & \textbf{DQN} & \textbf{SAC} \\
\midrule
Total Timesteps & $10^{7}$ & $10^{7}$ & $10^{7}$ \\
Learning Rate & $2.5 \times 10^{-4}$ & $2.5 \times 10^{-4}$ & $3 \times 10^{-4}$ \\
Discount Factor ($\gamma$) & 0.99 & 0.99 & 0.99 \\
Parallel Environments & 128 & 128 & 128 \\
Batch Size & $128 \times 128$ & 256 & 256 \\
GAE Lambda ($\lambda$) & 0.95 & N/A & N/A \\
Replay Buffer Size & N/A & $10^{6}$ & $10^{6}$ \\
Target Update Frequency & N/A & 2000 & 8000 \\
Entropy Coefficient ($\alpha$) & 0.01 & N/A & Auto-tuned \\
\bottomrule
\end{tabular}
\caption{Hyperparameters used for PPO, DQN, and SAC in the multi-agent test environments.}
\label{tab:hyperparams}
\end{table}

\subsection{Extensive Environment Performance Scaling}
\label{app:environment_scaling}

The following figures exhibit complete grid throughputs across varying observability modes on AMD Ryzen 9950x, AMD EPYC 7282, and the Intel 4-Core Laptop hardware profiles.

\bibliographystyle{plain}
\bibliography{references}

@article{bernstein2002complexity,
  title={The complexity of decentralized control of Markov decision processes},
  author={Bernstein, Daniel S and Givan, Robert and Immerman, Neil and Zilberstein, Shlomo},
  journal={Mathematics of operations research},
  volume={27},
  number={4},
  pages={819--840},
  year={2002},
  publisher={INFORMS}
}

@inproceedings{jin2020sample,
  title={Sample-efficient reinforcement learning of undercomplete {POMDPs}},
  author={Jin, Chi and Kakade, Sham and Krishnamurthy, Akshay and Liu, Qinghua},
  booktitle={Advances in Neural Information Processing Systems},
  volume={33},
  pages={18530--18539},
  year={2020}
}

@article{MARLreview,
  title={A review of cooperative multi-agent deep reinforcement learning},
  author={Oroojlooy, Afshin and Hajinezhad, Davood},
  journal={Applied Intelligence},
  volume={53},
  number={11},
  pages={13677--13722},
  year={2023},
  publisher={Springer}
}

@article{vinyals2019grandmaster,
  title={Grandmaster level in StarCraft II using multi-agent reinforcement learning},
  author={Vinyals, Oriol and Babuschkin, Igor and Czarnecki, Wojciech M and Mathieu, Micha{\"e}l and Dudzik, Andrew and Chung, Junyoung and Choi, David H and Powell, Richard and Ewalds, Timo and Georgiev, Petko and others},
  journal={nature},
  volume={575},
  number={7782},
  pages={350--354},
  year={2019},
  publisher={Nature Publishing Group UK London}
}

@article{berner2019dota,
  title={Dota 2 with large scale deep reinforcement learning},
  author={Berner, Christopher and Brockman, Greg and Chan, Brooke and Cheung, Vicki and D{\k{e}}biak, Przemys{\l}aw and Dennison, Christy and Farhi, David and Fischer, Quirin and Hashme, Shariq and Hesse, Chris and others},
  journal={arXiv preprint arXiv:1912.06680},
  year={2019}
}

@inproceedings{espeholt2018impala,
  title={Impala: Scalable distributed deep-rl with importance weighted actor-learner architectures},
  author={Espeholt, Lasse and Soyer, Hubert and Munos, Remi and Simonyan, Karen and Mnih, Vlad and Ward, Tom and Doron, Yotam and Firoiu, Vlad and Harley, Tim and Dunning, Iain and others},
  booktitle={International conference on machine learning},
  pages={1407--1416},
  year={2018},
  organization={PMLR}
}

@inproceedings{mnih2016asynchronous,
  title={Asynchronous methods for deep reinforcement learning},
  author={Mnih, Volodymyr and Badia, Adria Puigdomenech and Mirza, Mehdi and Graves, Alex and Lillicrap, Timothy and Harley, Tim and Silver, David and Kavukcuoglu, Koray},
  booktitle={International conference on machine learning},
  pages={1928--1937},
  year={2016},
  organization={PmLR}
}

@article{cusumano2025robust,
  title={Robust autonomy emerges from self-play},
  author={Cusumano-Towner, Marco and Hafner, David and Hertzberg, Alex and Huval, Brody and Petrenko, Aleksei and Vinitsky, Eugene and Wijmans, Erik and Killian, Taylor and Bowers, Stuart and Sener, Ozan and others},
  journal={arXiv preprint arXiv:2502.03349},
  year={2025}
}

@article{mittal2025isaac,
  title={Isaac lab: A gpu-accelerated simulation framework for multi-modal robot learning},
  author={Mittal, Mayank and Roth, Pascal and Tigue, James and Richard, Antoine and Zhang, Octi and Du, Peter and Serrano-Munoz, Antonio and Yao, Xinjie and Zurbr{\"u}gg, Ren{\'e} and Rudin, Nikita and others},
  journal={arXiv preprint arXiv:2511.04831},
  year={2025}
}

@article{weng2022envpool,
  title={Envpool: A highly parallel reinforcement learning environment execution engine},
  author={Weng, Jiayi and Lin, Min and Huang, Shengyi and Liu, Bo and Makoviichuk, Denys and Makoviychuk, Viktor and Liu, Zichen and Song, Yufan and Luo, Ting and Jiang, Yukun and others},
  journal={Advances in Neural Information Processing Systems},
  volume={35},
  pages={22409--22421},
  year={2022}
}

@article{torrellas1994false,
  title={False sharing and spatial locality in multiprocessor caches},
  author={Torrellas, Josep and Lam, Monica S and Hennessy, John L},
  journal={IEEE Transactions on Computers},
  volume={43},
  number={6},
  pages={651--663},
  year={1994},
  publisher={IEEE}
}

@article{arantes2025impact,
  title={Impact of Data-Oriented and Object-Oriented Design on Performance and Cache Utilization with Artificial Intelligence Algorithms in Multi-Threaded CPUs},
  author={Arantes, Gabriel M and Pinto, Richard F and Dalmazo, Bruno L and Borges, Eduardo N and Lucca, Giancarlo and de Mattos, Viviane LD and Cardoso, Fabian C and Berri, Rafael A},
  journal={arXiv preprint arXiv:2512.07841},
  year={2025}
}

@article{samvelyan2019starcraft,
  title={The starcraft multi-agent challenge},
  author={Samvelyan, Mikayel and Rashid, Tabish and De Witt, Christian Schroeder and Farquhar, Gregory and Nardelli, Nantas and Rudner, Tim GJ and Hung, Chia-Man and Torr, Philip HS and Foerster, Jakob and Whiteson, Shimon},
  journal={arXiv preprint arXiv:1902.04043},
  year={2019}
}

@article{ellis2023smacv2,
  title={Smacv2: An improved benchmark for cooperative multi-agent reinforcement learning},
  author={Ellis, Benjamin and Cook, Jonathan and Moalla, Skander and Samvelyan, Mikayel and Sun, Mingfei and Mahajan, Anuj and Foerster, Jakob and Whiteson, Shimon},
  journal={Advances in Neural Information Processing Systems},
  volume={36},
  pages={37567--37593},
  year={2023}
}

@article{terry2021pettingzoo,
  title={Pettingzoo: Gym for multi-agent reinforcement learning},
  author={Terry, J and Black, Benjamin and Grammel, Nathaniel and Jayakumar, Mario and Hari, Ananth and Sullivan, Ryan and Santos, Luis S and Dieffendahl, Clemens and Horsch, Caroline and Perez-Vicente, Rodrigo and others},
  journal={Advances in Neural Information Processing Systems},
  volume={34},
  pages={15032--15043},
  year={2021}
}

@article{suarez2024pufferlib,
  title={Pufferlib: Making reinforcement learning libraries and environments play nice},
  author={Suarez, Joseph},
  journal={arXiv preprint arXiv:2406.12905},
  year={2024}
}

@software{brax2021github,
  author = {C. Daniel Freeman and Erik Frey and Anton Raichuk and Sertan Girgin and Igor Mordatch and Olivier Bachem},
  title = {Brax - A Differentiable Physics Engine for Large Scale Rigid Body Simulation},
  url = {http://github.com/google/brax},
  version = {0.14.2},
  year = {2021},
}

@article{shacklett2023madrona,
  title={An Extensible, Data-Oriented Architecture for High-Performance, Many-World Simulation},
  author={Shacklett, Brennan and Zhan, Lucis and Chen, Hao and Sun, Mingfei and Fox, Dieter and Fatahalian, Kayvon},
  journal={ACM Transactions on Graphics (TOG)},
  volume={42},
  number={4},
  pages={1--14},
  year={2023},
  publisher={ACM New York, NY, USA}
}

@misc{ openmp08,
    author = {{OpenMP Architecture Review Board}},
    title = {{OpenMP} Application Program Interface Version 3.0},
    month = may,
    year = 2008,
    url = {http://www.openmp.org/mp-documents/spec30.pdf}
}

@inproceedings{gureya2020bandwidth,
  title={Bandwidth-aware page placement in numa},
  author={Gureya, David and Neto, Joao and Karimi, Reza and Barreto, Joao and Bhatotia, Pramod and Quema, Vivien and Rodrigues, Rodrigo and Romano, Paolo and Vlassov, Vladimir},
  booktitle={2020 IEEE International Parallel and Distributed Processing Symposium (IPDPS)},
  pages={546--556},
  year={2020},
  organization={IEEE}
}

@article{yu2022surprising,
  title={The surprising effectiveness of ppo in cooperative multi-agent games},
  author={Yu, Chao and Velu, Akash and Vinitsky, Eugene and Gao, Jiaxuan and Wang, Yu and Bayen, Alexandre and Wu, Yi},
  journal={Advances in neural information processing systems},
  volume={35},
  pages={24611--24624},
  year={2022}
}

@inproceedings{tavakoli2018action,
  title={Action branching architectures for deep reinforcement learning},
  author={Tavakoli, Arash and Pardo, Fabio and Kormushev, Petar},
  booktitle={Proceedings of the aaai conference on artificial intelligence},
  volume={32},
  number={1},
  year={2018}
}

@article{schulman2017proximal,
  title={Proximal policy optimization algorithms},
  author={Schulman, John and Wolski, Filip and Dhariwal, Prafulla and Radford, Alec and Klimov, Oleg},
  journal={arXiv preprint arXiv:1707.06347},
  year={2017}
}

@article{mnih2015human,
  title={Human-level control through deep reinforcement learning},
  author={Mnih, Volodymyr and Kavukcuoglu, Koray and Silver, David and Rusu, Andrei A and Veness, Joel and Bellemare, Marc G and Graves, Alex and Riedmiller, Martin and Fidjeland, Andreas K and Ostrovski, Georg and others},
  journal={nature},
  volume={518},
  number={7540},
  pages={529--533},
  year={2015},
  publisher={Nature Publishing Group}
}

@inproceedings{haarnoja2018soft,
  title={Soft actor-critic: Off-policy maximum entropy deep reinforcement learning with a stochastic actor},
  author={Haarnoja, Tuomas and Zhou, Aurick and Abbeel, Pieter and Levine, Sergey},
  booktitle={International conference on machine learning},
  pages={1861--1870},
  year={2018},
  organization={Pmlr}
}

@article{lanctot2019openspiel,
  title={OpenSpiel: A framework for reinforcement learning in games},
  author={Lanctot, Marc and Lockhart, Edward and Lespiau, Jean-Baptiste and Zambaldi, Vinicius and Upadhyay, Satyaki and P{\'e}rolat, Julien and Srinivasan, Sriram and Timbers, Finbarr and Tuyls, Karl and Omidshafiei, Shayegan and others},
  journal={arXiv preprint arXiv:1908.09453},
  year={2019}
}

@article{juliani2018unity,
  title={Unity: A general platform for intelligent agents},
  author={Juliani, Arthur and Berges, Vincent-Pierre and Teng, Ervin and Cohen, Andrew and Harper, Jonathan and Elion, Chris and Goy, Chris and Gao, Yuan and Henry, Hunter and Mattar, Marwan and others},
  journal={arXiv preprint arXiv:1809.02627},
  year={2018}
}

@article{towers2024gymnasium,
  title={Gymnasium: A standard interface for reinforcement learning environments},
  author={Towers, Mark and Kwiatkowski, Ariel and Terry, Jordan and Balis, John U and De Cola, Gianluca and Deleu, Tristan and Goul{\~a}o, Manuel and Kallinteris, Andreas and Krimmel, Markus and KG, Arjun and others},
  journal={arXiv preprint arXiv:2407.17032},
  year={2024}
}

@inproceedings{apex,
  title={Distributed prioritized experience replay},
  author={Dan, Horgan and Quan, J and Budden, D and others},
  booktitle={Proc. 5th Int. Conf. Learning Representations (ICLR, Vancouver, BC, Canada, 2018)},
  year={2018}
}

@article{pdqn,
  title={Simplifying deep temporal difference learning},
  author={Gallici, Matteo and Fellows, Mattie and Ellis, Benjamin and Pou, Bartomeu and Masmitja, Ivan and Foerster, Jakob Nicolaus and Martin, Mario},
  journal={arXiv preprint arXiv:2407.04811},
  year={2024}
}

@article{MinigridMiniworld23,
  author       = {Maxime Chevalier-Boisvert and Bolun Dai and Mark Towers and Rodrigo de Lazcano and Lucas Willems and Salem Lahlou and Suman Pal and Pablo Samuel Castro and Jordan Terry},
  title        = {Minigrid \& Miniworld: Modular \& Customizable Reinforcement Learning Environments for Goal-Oriented Tasks},
  journal      = {CoRR},
  volume       = {abs/2306.13831},
  year         = {2023},
}

@inproceedings{petrenko2020sample,
  title={Sample Factory: Egocentric 3D Control from Pixels at 100,000 FPS with a Single GPU},
  author={Petrenko, Aleksei and Huang, Zhehui and Kumar, Tushar and Sukhatme, Gaurav and Koltun, Vladlen},
  booktitle={International Conference on Machine Learning},
  pages={7654--7663},
  year={2020},
  organization={PMLR}
}

@manual{amdepyc7003hpctg,
  title        = {High Performance Computing ({HPC}) Tuning Guide for {AMD} {EPYC} 7003 Series Processors},
  author       = {{Advanced Micro Devices, Inc.}},
  organization = {Advanced Micro Devices, Inc.},
  year         = {2022},
  month        = {March},
  note         = {Document 70574; Revision 1.0},
  url          = {https://docs.amd.com/v/u/en-US/high-performance-computing-tuning-guide-amd-epyc7003-series-processors}
}

@misc{pybind11,
  author = {Wenzel Jakob and Jason Rhinelander and Dean Moldovan},
  year = {2017},
  note = {https://github.com/pybind/pybind11},
  title = {pybind11 -- Seamless operability between C++11 and Python}
}

@article{pytorch,
  title={Pytorch: An imperative style, high-performance deep learning library},
  author={Paszke, Adam and Gross, Sam and Massa, Francisco and Lerer, Adam and Bradbury, James and Chanan, Gregory and Killeen, Trevor and Lin, Zeming and Gimelshein, Natalia and Antiga, Luca and others},
  journal={Advances in neural information processing systems},
  volume={32},
  year={2019}
}

\end{document}